\documentclass[a4paper,aps,pra, reprint, amsmath,amsfonts,amssymb,superscriptaddress]{revtex4-2} 

\usepackage{dsfont}
\usepackage{graphicx}
\usepackage{physics}
\usepackage[version=4]{mhchem}
\usepackage{isotope}
\usepackage{bm}
\usepackage{hyperref}
\usepackage{tikz}
\usepackage{siunitx}
\usetikzlibrary{calc}
\usepackage[percent]{overpic}

\DeclareMathOperator{\erfi}{\mathrm{erfi}}
\newcommand{\dyarr}[1]{\overleftrightarrow{#1}}
\begin{document}

\title{Superradiance and anomalous hyperfine splitting in inhomogeneous ensembles}

\begin{abstract}
Collective effects in the interaction of light with ensembles of identical scatterers play an important role in many fields of physics. However, often the term ``identical'' is not accurate due to the presence of  hyperfine fields which induce inhomogeneous transition shifts and splittings. Here we develop a formalism based on the Green function method to model the linear response of such inhomogeneous ensembles in one-dimensional waveguides. We obtain a compact  formula for the collective spectrum, which exhibits deviations from the uniform frequency shift and broadening expected of two level systems. In particular, if the coherent contribution to the collective coupling is large, the effect of inhomogeneous broadening can be suppressed, with the linewidth approaching that of the superradiant value. We apply this formalism to describe collective effects in x-ray scattering off thin-film waveguides for inhomogeneous hyperfine parameters. 

\end{abstract}

%
%
%
%
\author{Petar Andreji{\'c}}
\email{petar.andrejic@mpi-hd.mpg.de}
\affiliation{Max Planck Institut f{\"ur} Kernphysik, Saupfercheckweg 1, D-69117 Heidelberg, Germany}
\author{Adriana P{\' a}lffy}
\email{adriana.palffy-buss@fau.de}
\affiliation{Max Planck Institut f{\"ur} Kernphysik, Saupfercheckweg 1, D-69117 Heidelberg, Germany}
\affiliation{Department of Physics, Friedrich-Alexander-Universit\"at Erlangen-N\"urnberg,  D-91058 Erlangen, Germany}

\date{\today}
\maketitle

\section{Introduction}
When an ensemble of identical atoms interacts with light of wavelength much larger than the size of the ensemble, the atoms absorb and emit radiation collectively, resulting in the phenomenon of superradiance. This  has been first investigated theoretically by Dicke for the case of two-level systems in 1954~\cite{dickeCoherenceSpontaneousRadiation1954}. Provided  the average inter-particle distance is much smaller than the average radiated wavelength, the permutational invariance of the collective light-matter interaction results in the $N$ atoms of the ensemble absorbing and emitting radiation collectively, with a factor of $N$ enhancement of the decay rate compared to the single atom value  \cite{dickeCoherenceSpontaneousRadiation1954,Rehler1971,Haroche1982,Leonardi1986review}. Following the pioneering experiment by  Skribanowitz and co-workers \cite{Skribanowitz1973},  superradiance has been demonstrated and studied in many systems, such as Bose-Einstein condensates \cite{Inouye1999,Baumann2010}, quantum dots \cite{Scheibner2007}, color centers \cite{Angerer2018},  cold atoms \cite{araujoSuperradianceLargeDilute2016a,cottierRoleDisorderSuper2018a}, M\"ossbauer nuclei 
\cite{xrayReview2013,rohlsbergerCollectiveLambShift2010,Chumakov2017}  and trapped atoms coupled to a cavity \cite{Reimann2015,Kim2018}, to name a few. Thus, the concept of superradiance has implications in many fields, such as quantum information  \cite{scullyDirectedSpontaneousEmission2006,deoliveiraSinglephotonSuperradianceCold2014}, cavity quantum electrodynamics \cite{Reimann2015,Kim2018}, astrophysics \cite{Rajabi2016} or advanced light source  \cite{Vieira2021}. 

The mechanism behind superradiance and collective emission also applies to multilevel systems, and extensions to the Dicke model for multilevel atoms have been explored~\cite{arecchiCoherentStatesRlevel1973,mullerMultipleScatteringPhotons2005a,Gegg2016,sutherlandSuperradianceInvertedMultilevel2017,geggIdenticalEmittersCollective2017,kongCollectiveRadiationSpectrum2017,SumantaI-2018}. Like the Dicke model, these assume completely uniform illumination of the ensemble of atoms, and that the latter are all completely identical. In parallel, since the 1970s theoretical works have been addressing the effect of inhomogeneous broadening on superradiance \cite{Agarwal1971,Eberly1971,Eberly1971-Acta,Jodoin1974,Leonardi1982,MolAggr1989,Temnov2005}.  This is pertinent to many physical systems where the energy levels and decay rates are different due to local environment effects such as Zeeman and hyperfine splittings induced by  magnetic or electric fields, Doppler broadening due to thermal interactions, or simply by fabrication in the case of quantum dots. Inhomogeneous broadening effects have been also investigated in the related process of superfluorescence \cite{Bonifacio1975,Haake1980,Ishikawa2016}. It is the purpose of this work to present a versatile formalism which allows to theoretically model  the superradiant response of multilevel systems in one-dimensional waveguides, where in the same time the emitters are inhomogeneous. 

One-dimensional waveguides play a special role for superradiance because they facilitate the otherwise challenging uniform illumination of the scatterer ensemble.  Superradiance relies on permutation invariance, i.e., the invariance of the system under the exchange of any two scatterers in the ensemble, which in turn relies on uniform illumination of the ensemble.  In the single-photon regime, incident pulses are re-emitted in a highly directional manner~\cite{scullyDirectedSpontaneousEmission2006, araujoSuperradianceLargeDilute2016a}, preserving the incident wave-vector. As such, it has been found that arranging the atoms in quasi one dimensional arrangements such as pencil geometries can enhance  superradiance~\cite{roofObservationSinglePhotonSuperradiance2016}. Also, uniform illumination can be easily achieved if the atoms are placed in a one dimensional waveguide like structure~\cite{ruostekoskiEmergenceCorrelatedOptics2016a, asenjo-garciaAtomlightInteractionsQuasionedimensional2017,SumantaII-2018}, or x-ray grazing incidence reflection from thin films~\cite{benedictCoherentReflectionSuperradiation1988,samsonInducedSuperradianceThin1990,limExcitonDelocalizationSuperradiance2004,heegCollectiveEffectsMultiple2015,lentrodtInitioQuantumModels2020}. The atomic excitations propagate through the waveguide as a polariton, and the waveguide structure restricts the propagation of the scattered light to one dimensional, plane wave propagation. This results in uniform illumination, with translational symmetry playing the role of permutational symmetry, achieving superradiance without requiring the wavelength be much larger than the atomic spacing. 

In this work, we investigate the superradiant response of multilevel and inhomogeneous scatterers, e.g., atoms or nuclei, in one-dimensional waveguides.  If the length scale of the environmental variation is much larger than the emitter spacing, the ensemble can be partitioned into approximately uniform sub-ensembles. Thus, the collective interaction can still play a significant role, however the interplay between the spectral inhomogeneity and the collective scattering results in non-trivial structure of the resulting spectra. To describe the dipole-dipole interaction we use a Green's function method, developed by Gr{\" u}ner and Welsch~\cite{grunerGreenfunctionApproachRadiationfield1996}, which has been successfully applied to describe superradiance in diverse systems such as atomic clouds~\cite{cottierRoleDisorderSuper2018a, svidzinskyCooperativeSpontaneousEmission2008, svidzinskyCooperativeSpontaneousEmission2010, araujoSuperradianceLargeDilute2016a}, one-dimensional waveguides~\cite{ruostekoskiEmergenceCorrelatedOptics2016a,asenjo-garciaAtomlightInteractionsQuasionedimensional2017}, and thin film x-ray reflection ~\cite{lentrodtInitioQuantumModels2020, kongGreenSfunctionFormalism2020}. A compact formula is found for the weak-excitation regime susceptibility, in terms of the coherent average of the emitters responses, with the collective interaction describable via a single complex constant. This allows for the effects of inhomogeneities and of the collective interaction to be analysed separately, allowing for a better understanding of their respective contributions to the collective spectrum.

We apply this formalism to the concrete example of x-ray quantum optics systems that comprise of ensembles of M\"ossbauer nuclei in thin-film x-ray cavities. 
The latter are a particularly suitable platform for exploring superradiance and collective interaction between emitters. In these systems, a thin layer of resonant nuclei is placed in the centre of a thin-film cavity, forming a waveguide like structure. Evanescent guided modes of the cavity are driven at grazing incidence, coupling to M{\" o}ssbauer transitions in the nuclei. This forms an exceptionally `clean' system, due to the incredibly narrow linewidth of M{\" o}ssbauer transitions, and the large energy scale of the x-rays resulting in effectively zero thermal noise~\cite{xrayReview2013}. Using this set-up, superradiance of single x-ray photons was experimentally demonstrated by R{\" o}hlsberger \textit{et al}~\cite{rohlsbergerCollectiveLambShift2010}, observing both a collective Lamb shift, and up to a factor of 61 enhancement of the decay rate. Further work by Chumakov \textit{et al} at the SACLA x-ray free electron laser has verified the scaling behaviour for multi-photon excitations~\cite{Chumakov2017}. 

However, in the same time  the resonant nuclear spectrum often features hyperfine interactions due to the electronic and magnetic environment. As creating large single crystal samples is challenging, the resonant layer will in general be poly-crystalline, and therefore the nuclear hyperfine environment is typically inhomogeneous. Our results consider  examples of the combination of collective and inhomogeneous effects, and show that in general the addition of collective interactions is more complex than a simple broadening and Lamb shift. In particular, the coherent part of the collective coupling distorts the line-shape, and can even be used to reduce the broadening from the sample inhomogeneity, thus providing user control over the samples linewidth. These findings are therefore very useful both for understanding experimental results, as well as for designing schemes to control the collective radiation spectrum and remove unwanted features. Due to the versatility of the formalism,  these results will be useful for a larger community investigating superradiant effects in various types of one-dimensional ensembles.

This work is structured as follows. Section \ref{sec:model} introduces the general model and derives the equation of motion and the collective Lamb shift and cross couplings. The application of the model to x-ray  thin-film cavities with embedded layers of M\"ossbauer nuclei is presented in Sec.~\ref{sec:xqo}. On this occasion we put the present formalism in the context of already existing theoretical models for x-ray grazing incidence on thin-film cavities in Sec.~\ref{sec:history}. 
 Our examples for inhomogeneous nuclear hyperfine splitting are discussed in Sec.~\ref{sec:example1} and \ref{sec:example2}. Conclusions and a brief outlook are summarized in Sec.~\ref{sec:concl}.








\section{Model}\label{sec:model}
\subsection{Hamiltonian and Lindblad operators}

\begin{figure}
  \includegraphics[width=8.6 cm]{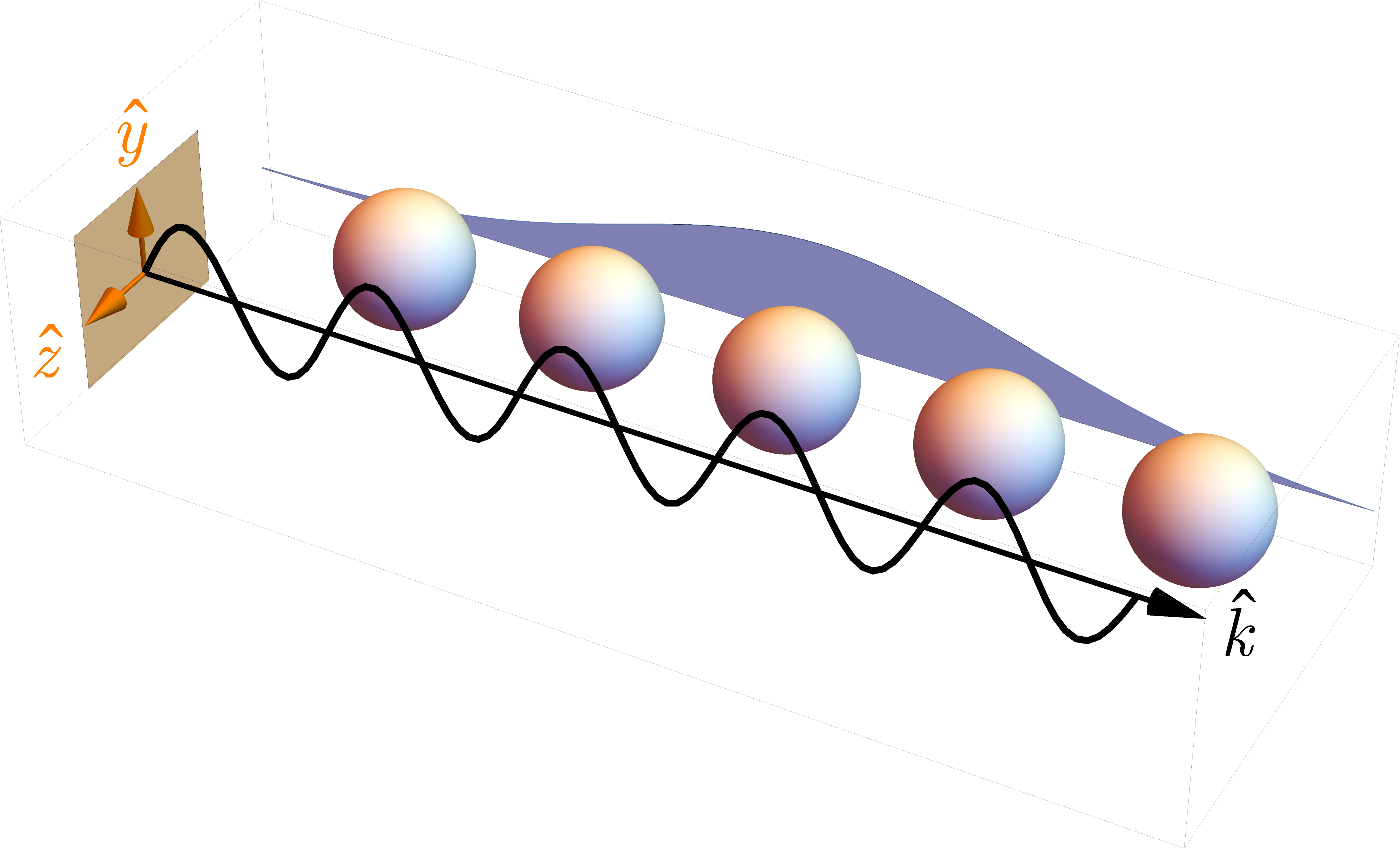}
  \caption{\label{fig:atom_plot} Schematic plot (not to scale) of waveguide scattering geometry. The scatterers (spheres) are lined up along the waveguide direction $\hat{k}=\hat{x}$. The restriction of propagation to this one dimension requires the polarisation of the guided modes to span the plane $\hat{y},\hat{z}$. As an example, we may consider a superimposed magnetic hyperfine field distribution with Gaussian profile (blue shaded curve in the background), creating an inhomogeneity.}
\end{figure}

We begin with an ensemble of atoms in a one-dimensional waveguide, schematically illustrated in Fig.~\ref{fig:atom_plot}. The spectral parameters of the atoms are inhomogeneous, and the inhomogeneity is assumed to vary slowly over the inter-atomic length scale, such that the atoms can be divided into equally sized sub-ensembles that are approximately translationally and permutationally symmetric. The size of each sub-ensemble, and hence the number required, is determined by the gradient of the inhomogeneity over the spatial extent of the atoms. The variation of the inhomogeneity across each sub-ensemble should be taken to be small enough that it cannot be resolved within the linewidth of the transitions present, and can therefore be treated as a negligible perturbation. The atoms are driven by a probe field $  \vb{E}_{\text{p}}(t)e^{i(k_0 x-\omega_0 t)}$ of frequency $\omega_0$, wave vector $k_0$, and uniform illumination, with a possible time dependent envelope.

Following~\cite{lentrodtInitioQuantumModels2020,asenjo-garciaAtomlightInteractionsQuasionedimensional2017}, we work in the rotating frame of the driving field. The internal Hamiltonian for the atoms is given by
\begin{equation}
  H_{\text{A}} = -\sum_n \sum_{i\in\mathcal{D}_n}\sum_{\mu \in \mathcal{T}_n} \hbar\Delta_{\mu} \ketbra{e_{\mu}^{(i)}}{e_{\mu}^{(i)}},
\end{equation}
where $\mathcal{D}_n$ is the set of atoms in sub-ensemble $n$, $\mathcal{T}_n$ is the set of excited states of sub-ensemble $n$, and $\Delta_{\mu}\ll \omega_0$ is the detuning of excited state $\mu$. Furthermore, $\hbar$ is the reduced Planck constant.

Incoherent decay is described by the Lindblad operator
\begin{equation}
  L_{\text{A}}[\rho] = -\sum_n \sum_{i\in\mathcal{D}_n}\sum_{\mu \in \mathcal{T}_n}
    \hbar \gamma_{\mu} \mathcal{L}\left[\rho, \ketbra{e_{\mu}^{(i)}}{g^{(i)}}, \ketbra{g^{(i)}}{e_{\mu}^{(i)}}\right].
\end{equation}
where $\gamma_\mu$ is the natural decay rate of excited state $\mu$, and
\begin{equation}
  \mathcal{L}[\rho,A,B] = AB\rho + \rho AB - 2B\rho A.
\end{equation}
At low saturations, the probe field only drives dipole transitions directly accessible from the ground state. It will be convenient to express the transition dipole vectors in the form $\vb{d}_\mu \wp$, with $\vb{d}_\mu$ a dimensionless vector, and $\wp$ the mean dipole magnitude.

The driving Hamiltonian is then given by
\begin{equation}
  H_{\text{p}} = \wp\sum_{n} \sum_{i\in\mathcal{D}_n}\sum_{\mu \in \mathcal{T}_n}  \vb{d}_{\mu} \cdot \vb{E}_{\text{p}}(t) e^{ik_0 x_{i}} \ketbra{e_\mu^{(i)}}{g^{(i)}} + \text{h.c.}
\end{equation}

The atoms couple collectively via a dipole-dipole interaction, described by the Green's function formalism~\cite{lentrodtInitioQuantumModels2020,asenjo-garciaAtomlightInteractionsQuasionedimensional2017,grunerGreenfunctionApproachRadiationfield1996}. This gives the matrix elements for the dipole-dipole interaction via the classical dyadic Green's function for the waveguide,
\begin{widetext}
\begin{equation}\label{eq:dipole-dipole}
\begin{aligned}
  H_{\text{dd}} &= 
  -\mu_0 \omega_0^2 \wp^2 
  \sum_{n,m}
  \sum_{\substack{i \in \mathcal{D}_{n},\\ j \in \mathcal{D}_m}} 
  \sum_{\substack{\mu \in \mathcal{T}_n,\\ \nu \in \mathcal{T}_m}}  
  \vb{d}_{\mu} \cdot \Re (\dyarr{G}(x_{i}, x_{j})) \cdot \vb{d}_{\nu}^* 
  \ketbra{e_{\mu}^{(i)}}{g^{(i)}} \otimes \ketbra{g^{(j)}}{e_{\nu}^{(j)}}
  + \text{h.c.}
  \\
  L_{\text{dd}}[\rho] &= -\mu_0 \omega_0^2 \wp^2 
  \sum_{n,m}
  \sum_{\substack{i \in \mathcal{D}_{n},\\ j \in \mathcal{D}_m}} 
  \sum_{\substack{\mu \in \mathcal{T}_n,\\ \nu \in \mathcal{T}_m}}  
  \vb{d}_{\mu} \cdot \Im (\dyarr{G}(x_{i}, x_{j})) \cdot \vb{d}_{\nu}^* 
  \mathcal{L}\left[\rho,\ketbra{e_{\mu}^{(i)}}{g^{(i)}},\ketbra{g^{(j)}}{e_{\nu}^{(j)}}\right],
\end{aligned}
\end{equation}
\end{widetext}
where $\dyarr{G}(x_i, x_j)$ is the dyadic Green's function for the waveguide, and $\mu_0$  the vacuum permeability, respectively.

Due to the one-dimensional propagation, and translational symmetry, the dyadic Green's function $\dyarr{G}(x_i, x_j, \omega)$ can be expressed in the form
\begin{equation}
  \dyarr{G}(x_i, x_j, \omega) = \dyarr{\mathds{1}}^\perp \int \frac{\dd{k}}{2\pi L^{-1}} \mathcal{G}(k, \omega) e^{i k (x_i - x_j)},
\end{equation}
where $\mathcal{G}(k)$ is a scalar function, $\dyarr{\mathds{1}}^\perp$ is a projection matrix for the 2D subspace perpendicular to the guided direction (Figure~\ref{fig:atom_plot}), and $L$ is the quantisation length of the ensemble.

Following~\cite{lentrodtInitioQuantumModels2020}, we make the approximation, valid at low saturation, that the scattered radiation always has the same wave-vector as the driving field. Therefore, for the purposes of Eq.~\eqref{eq:dipole-dipole}, we can substitute 
\begin{equation}
\dyarr{G}(x_i, x_j,\omega_0) \to \dyarr{\mathds{1}}^\perp \mathcal{G}(k_0, \omega_0)e^{ik_0(x_i - x_j)}.
\end{equation}
The plane wave phase factors can then be eliminated by the following unitary transformation:
\begin{equation}
  \ket{e_{\nu}^{(j)}}\to e^{-ik_0 x_{j}}\ket{e_{\nu}^{(j)}}.
\end{equation}
The model then becomes permutationally symmetric. Such systems can be analysed by a  generalisation of the Holstein-Primakoff transformation~\cite{kuruczMultilevelHolsteinPrimakoffApproximation2010}. This maps the collective transitions of the system to independent Bosonic modes, with the collective ground state mapping to the Bosonic vacuum. We introduce the bosonic creation (annihilation) operators $b^\dagger_\mu$ ($b_\mu$) for each collective transition $\mu$. For our system, the transformation reads
\begin{equation}
\begin{aligned}
  \sum_{i\in \mathcal{D}_n}\ketbra{e_{\mu}^{(i)}}{e_{\nu}^{(i)}} &= b_{\mu}^{\dagger}b_{\nu},
  \quad  \mu,\nu \in \mathcal{T}_n, 
  \\
  \sum_{i \in \mathcal{D}_n} e^{ik_0 x_{i}}\ketbra{e_{\mu}^{(i)}}{g^{(i)}}& = b_{\mu}^\dagger\sqrt{N p_n - \sum_{\nu \in \mathcal{T}_n}  b_{\nu}^\dagger b_{\nu}} 
  \\
  &\approx \sqrt{N p_n} b_{\mu}^\dagger, \quad \mu \in \mathcal{T}_n,
\end{aligned}
\end{equation}
where $p_n$ is the proportion of atoms in sub-ensemble $n$, $N$ is the total number of atoms in all ensembles, and
\begin{equation}
  [b_\nu, b_\mu^\dagger] = \delta_{\mu\nu}\delta_{mn}, \quad \mu \in \mathcal{T}_n, \nu \in \mathcal{T}_m.
\end{equation}
As we are interested in the linear response of our system, we have $\expval{b_\mu^\dagger b_\mu}\ll \sqrt{N}$, and we may linearise the collective transition operators,
\begin{equation}
	\sum_{i \in \mathcal{D}_n} e^{ik_0 x_{i}}\ketbra{e_{\mu}^{(i)}}{g^{(i)}} \approx \sqrt{N p_n} b_{\mu}^\dagger, \quad \mu \in \mathcal{T}_n.
\end{equation}
In terms of these Bosonic operators, the Hamiltonian of the linearised system reads
\begin{equation}\label{eq:final-hamiltonian}
\begin{aligned}
  H_{\text{A}} &= -\sum_n\sum_{\mu \in \mathcal{T}_n} \hbar \Delta_{\mu} \, b_{\mu}^\dagger b_{\mu},
  \\
  H_{\text{dd}} &= -\hbar J 
  \sum_{n,m}
  \sum_{\substack{\mu\in\mathcal{T}_n,\\ \nu\in\mathcal{T}_m}}
  \sqrt{p_n p_m} \vb{d}^\perp_\mu \cdot \vb{d}_\nu^{*\perp} \,b_\mu^\dagger b_\nu,
  \\
  H_{\text{p}} &=  \sqrt{N}\wp 
  \sum_n
  \sum_{\mu\in\mathcal{T}_n}
  \sqrt{p_n}\vb{d}_\mu^\perp \cdot \vb{E}_{\text{p}}(t)\, b_\mu^\dagger + \text{h.c.},
  \\
  L_{\text{A}}[\rho] &= -\sum_n
  \sum_{\mu\in\mathcal{T}_n} 
    \hbar \gamma_\mu \mathcal{L}[\rho, b_\mu^\dagger, b_\mu],
  \\
  L_{\text{dd}}[\rho] &= -\hbar \Gamma\
  \sum_{n,m}
  \sum_{\substack{\mu\in\mathcal{T}_n,\\ \nu\in\mathcal{T}_m}} \sqrt{p_n p_m}\vb{d}^\perp_\mu \cdot \vb{d}_\nu^{*\perp}
    \mathcal{L}[\rho,b_\mu^\dagger, b_\nu],
\end{aligned}
\end{equation}
where
\begin{equation}\label{eq:j-gamma-from-g}
J +i\Gamma = N\frac{\mu_0 \omega_0^2 \wp^2}{\hbar}\mathcal{G}(k_0, \omega_0), 
\quad 
\vb{d}_\mu^{\perp}= \vb{d}_\mu \cdot \dyarr{\mathds{1}}^\perp.
\end{equation}
Note that in transforming $L_{\text{A}}$ to the Bosonic operators, we have assumed that the decay is dominated by the superradiance, $\Gamma\gg \gamma_\mu$, such that we can approximate the single particle decay as collective. A more exact treatment of the single particle decay is given by Shammah \textit{et al.}~\cite{shammahOpenQuantumSystems2018}, and Gegg~\cite{Gegg2016,geggIdenticalEmittersCollective2017}, but is not significant in the low saturation regime we are considering (see Appendix~\ref{app:collective-decay}). 

Finally, although we have considered only a single ground state, for systems with multiple ground states the resulting equations of motion are of the same form. The partitioning of the sub-ensembles can be extended to partition by the initial ground state configurations of the atoms. Terms that couple to a different ground state configuration are suppressed by a factor of $\sqrt{N}$, and can be neglected (see Appendix~\ref{app:multiple-ground}).

\subsection{Equation of motion}
The equation of motion for a transition operator corresponding to $\mu \in \mathcal{T}_n$ is given by
\begin{equation}\label{eq:equation-of-motion}
\begin{aligned}
  \partial_t b_\mu &= \frac{i}{\hbar}[H, b_\mu] +\frac{1}{\hbar}L[b_\mu],
  \\
  &= (i\Delta_\mu-\gamma_\mu)b_\mu 
  \\
  &\qquad + (i J - \Gamma) 
  \sqrt{p_n} \vb{d}^\perp_\mu \cdot \sum_m \sqrt{p_m}\sum_{\nu\in T_m} \vb{d}_\nu^{*\perp} b_\nu 
  \\
  &\qquad- i \frac{\sqrt{N}\wp}{\hbar}\sqrt{p_n}\vb{d}_\mu^\perp \cdot \vb{E}_{\text{p}}(t).
\end{aligned}
\end{equation}

Due to the one dimensional nature of the problem, observables will depend on the dipole response only in the two dimensional polarisation space of the guided modes. As such, we need only consider the transverse polarisation operator,
\begin{equation}
\begin{aligned}
  \vb{P} &= \wp \sum_{n} \sum_{\mu \in \mathcal{T}_n}\sum_{i\in \mathcal{D}_n} \ketbra{g^{(i)}}{e_\mu^{(i)}}\vb{d}_\mu^\perp 
  \\
  &\approx \sqrt{N} \wp \sum_n \sqrt{p_n} \sum_{\mu\in \mathcal{T}_n}\vb{d}_\mu^{*\perp}  b_\mu.
\end{aligned}
\end{equation}
The susceptibility is found via solving for the linear response,
\begin{equation}
  \vb{P}(\omega) = \epsilon_0 \dyarr{\chi}(\omega) \cdot\vb{E}_{\text{p}}(\omega),
\end{equation}
where $\epsilon_0$ is the vacuum permittivity.
A compact solution for this can be obtained from the equation of motion, in Fourier space. We first Fourier transform Equation~\eqref{eq:equation-of-motion} to obtain
\begin{widetext}
\begin{equation}\label{eq:fourier-eom}
  (\omega + \Delta_\mu+i\gamma_\mu)b_\mu(\omega)
  + 
  (J+i\Gamma) \sqrt{p_n} \vb{d}^\perp_\mu 
  \cdot 
  \sum_m \sqrt{p_m}\sum_{\nu\in \mathcal{T}_m}
  \vb{d}_\nu^{*\perp} b_\nu(\omega) 
  =
  \frac{\sqrt{N}\wp}{\hbar} \sqrt{p_n}\vb{d}_\mu^\perp \cdot\vb{E}_{\text{p}}(\omega).
\end{equation}
\end{widetext}
It will be useful at this stage to define a reference frequency scale $\gamma_0$. For example, this could be the natural linewidth of a single atom, which for an electric dipole excitation can be expressed in terms of the mean dipole magnitude as $\omega_0^3 \wp^2 {(3\pi \hbar\epsilon_0 c^3)}^{-1}$. 

Multiplying both sides of~\eqref{eq:fourier-eom} by
\begin{equation}
	\frac{\sqrt{N p_n} \wp \vb{d}_\mu^{*\perp}}{\omega + \Delta_\mu + i\gamma_\mu},
\end{equation}
and summing over $n,\mu\in \mathcal{T}_n$, we obtain
\begin{equation}
  \vb{P}(\omega) + G \dyarr{\mathcal{F}}(\omega) \cdot \vb{P}(\omega) = \frac{N \wp^2}{\hbar\gamma_0} \dyarr{\mathcal{F}}(\omega) \cdot \vb{E}_{\text{p}}(\omega),
\end{equation}
where $G = \gamma_0^{-1}(J+i\Gamma)$, and
\begin{equation}\label{eq:single-particle-response}
  \dyarr{\mathcal{F}}(\omega) = \sum_n  p_n \sum_{\mu\in \mathcal{T}_n} \frac{\gamma_0 \vb{d}_\mu^{*\perp} \otimes \vb{d}_\mu^\perp }{\omega + \Delta_\mu +i\gamma_\mu}
\end{equation}
is the `layer response matrix', the coherent average of the responses of each transition in each sub-ensemble. Note that in this context $\otimes$ refers to the outer product. Such a quantity appears in the dynamical scattering approach to x-ray propagation~\cite[Eq.~(4.13)]{rohlsbergerNuclearCondensedMatter2004},~\cite{hannonResonantExchange1988}.

Solving for the polarisation we obtain
\begin{equation}
  \vb{P}(\omega) = \frac{N \wp^2}{\hbar\gamma_0} {\left(\dyarr{\mathds{1}} + G\dyarr{\mathcal{F}}(\omega)\right)}^{-1}  \cdot \dyarr{\mathcal{F}}(\omega) \cdot \vb{E}_{\text{p}}(\omega).
\end{equation}
This directly gives the first order susceptibility,
\begin{equation}\label{eq:collective-response}
  \dyarr{\chi}(\omega) = \chi_0 {\left(\dyarr{\mathds{1}} + G\dyarr{\mathcal{F}}(\omega)\right)}^{-1} \cdot \dyarr{\mathcal{F}}(\omega),
\end{equation}
where $\chi_0 = N \wp^2{(\hbar\gamma_0 \epsilon_0)}^{-1}$.

\subsection{Collective Lamb shift and cross-couplings}
The coherent part of the collective coupling $J$ has previously been referred to as a collective Lamb shift~\cite{ScullyLambShift2010,rohlsbergerCollectiveLambShift2010,wen2019}. Indeed, in the limit of a single, uniform transition with natural decay width $\gamma_0$, the susceptibility can be treated as a scalar, and is given by
\begin{equation}
  \chi(\omega) =\chi_0 \frac{\gamma_0}{\omega + \Delta +i \gamma + J + i \Gamma},
\end{equation}
describing a single line, shifted by $J$ and broadened by $\Gamma$. 
However, with multilevel atoms, and inhomogeneous configurations, $J$ does not act just as a Lamb shift, but also provides additional cross-couplings, analogous to an additional control field between transitions.

Specifically, a Lamb shift is a shift in an energy level due to emission and re-absorption of virtual photons from the same state. The analogue of this in Equation~\eqref{eq:final-hamiltonian} is given by the diagonal matrix elements of $H_{\text{dd}}$,
\begin{equation}
-\hbar J \vb{d}_\mu^\perp \cdot \vb{d}_\mu^{*\perp}.
\end{equation}

If the collective coupling acts purely as a `Lamb shift', then each transition is simply shifted and broadened, giving a susceptibility of
\begin{equation}\label{eq:wrong-lamb-shift}
\overleftarrow{\chi}(\omega) = \chi_0 \sum_\mu \frac{\gamma_0 \vb{d}_\mu^{*\perp} \otimes \vb{d}_\mu^\perp}{\omega + \Delta_\mu + i\gamma_\mu + (J+i\Gamma)\vb{d}^\perp_\mu \cdot \vb{d}_\mu^{*\perp}}.
\end{equation}
However, we can see in Eq.~\eqref{eq:final-hamiltonian} that due to additional cross-couplings from $J$, and spontaneously generated coherences from $\Gamma$ \cite{heegXrayQuantumOptics2013},  the resulting spectrum will  not be so straightforward to interpret, and is more generally described by Eq.~\eqref{eq:collective-response}. Indeed, as we shall see in Sec.~\ref{sec:example1} and \ref{sec:example2}, the spectrum for inhomogeneous two level atoms shows features that cannot be attributed to a simple Lamb shift. In the context of x-ray quantum optics, this was already hinted at in Ref.~\cite{heegXrayQuantumOptics2013} and later on discussed in more detail for the general case in Ref.~\cite{kongCollectiveRadiationSpectrum2017}.
We note that in particular, Ref.~\cite{kongCollectiveRadiationSpectrum2017} has addressed the specific case of uniform magnetic hyperfine splitting, using  the atomic cloud model of Svidzinsky~\textit{et al}~\cite{svidzinskyCooperativeSpontaneousEmission2010} to describe the collective coupling. This method derives an interaction kernel for the effective inter-atomic interactions, which turns out to be identical to the free space Green's function. As such, the Hamiltonian of our model reproduces the model of Ref.~\cite{kongCollectiveRadiationSpectrum2017} in the limiting case of atoms in free space, and uniform magnetic splitting.  

\section{Application to x-ray quantum optics \label{sec:xqo}}

\begin{figure}
\begin{overpic}[width=8.6 cm]{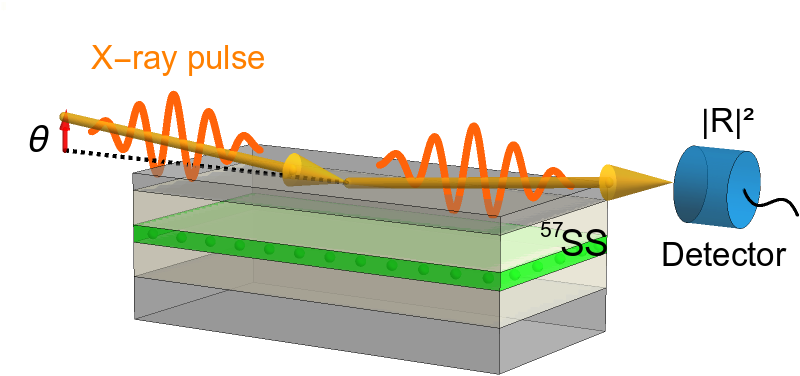}
\put (84,40) {(a)}
\end{overpic}

\vspace{0.15cm}

\begin{overpic}[width=8.0 cm]{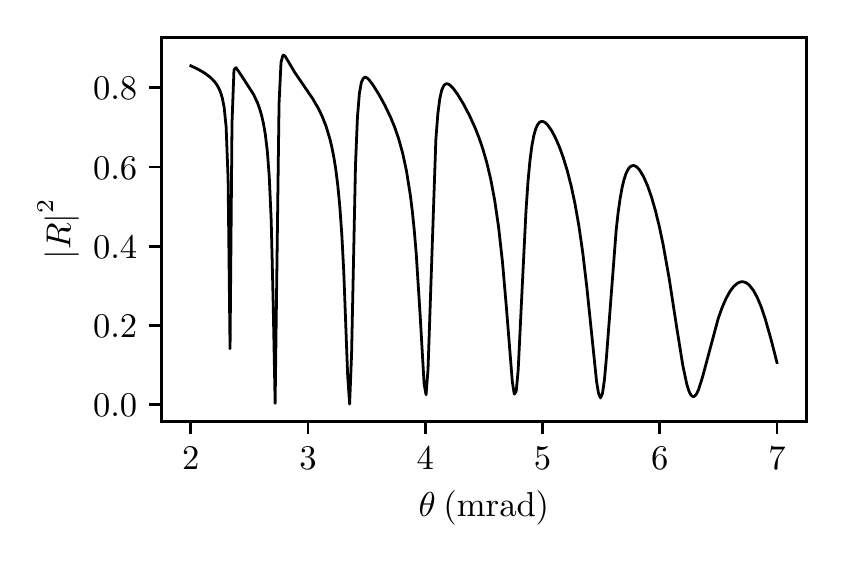}
\put (86.8,56) {(b)}
\end{overpic}
\caption{\label{fig:cavity}(a) Schematic illustration of a thin-film cavity. x-rays in grazing incidence with angle $\theta$ couple evanescently to the layered structure, exciting resonant transitions in stainless steel 95\% enriched with \isotope[57]{Fe} (referred to as \isotope[57]{SS}). The cavity reflectivity $|R|^2$ is measured at the detector. 
\\
(b) Example of theoretical reflectivity spectrum $|R|^2(\theta)$ for a cavity with structure  Pt \SI{2.8}{\nano\metre}/C \SI{22}{\nano\metre}/\mbox{\isotope[57]{SS} \SI{0.6}{\nano\metre}}/ \mbox{C \SI{22.5}{\nano\metre}}/Pt \SI{15}{\nano\metre}, obtained using the Python library {\it pynuss} \cite{pynuss}.  The reflectivity has various minima, that correspond to the resonant guided modes.}
\end{figure}

We now discuss the application of the general model from Section~\ref{sec:model} to x-ray quantum optics with M{\" o}ssbauer nuclei, and the connection to existing formalisms. The relevant experimental setup comprises so-called x-ray thin-film cavities using grazing incidence reflection as illustrated in Fig.~\ref{fig:cavity}(a).   
In this setup, layers are stacked from alternating high and low atomic number $Z$ materials (for instance, Pt or Pd alternating with C or \ce{B4C})  to form a waveguide structure for a pulse fired at grazing incidence to the layers. A thin resonant layer of M\"ossbauer nuclei, for instance \isotope[57]{Fe} or \isotope[57]{Fe}-enriched stainess steel (\isotope[57]{SS}) is embedded in this stack, usually sandwiched between low-$Z$ material layers. 
The scattering response of the system is recorded in the cavity reflectivity measured at the detector. As a function of incidence angle, the cavity reflectivity presents several minima, which correspond to the resonant driving of guided modes. An example is presented in Fig.~\ref{fig:cavity}(b). The minima are known as critical angles and they indicate the formation of a standing wave structure across the cavity layers. The resonant layer of M\"ossbauer nuclei is made sufficiently thin such that the guided mode field is approximately uniform across the depth of the layer.

\begin{figure}
\includegraphics[width=8.6cm]{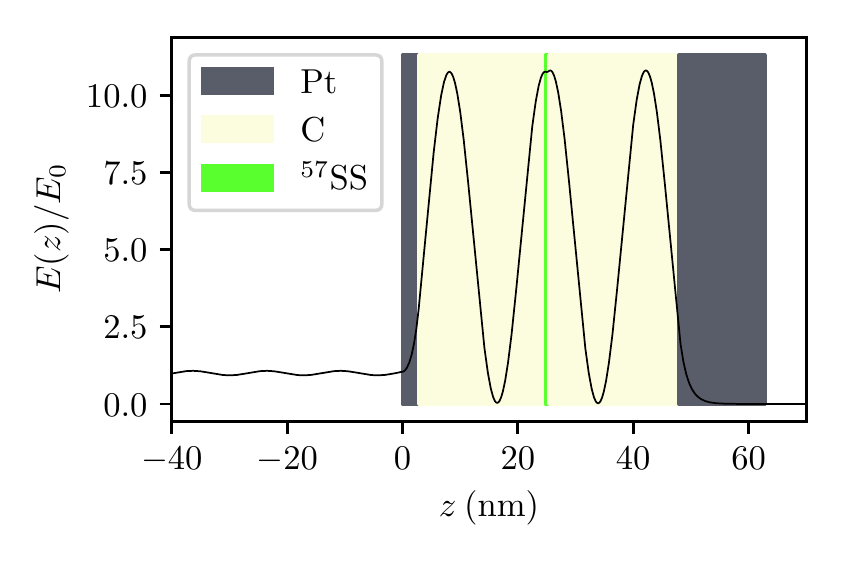}
\caption{\label{fig:field} Probe field intensity profile throughout the sample depth $z$, for the third reflection minimum in Figure~\ref{fig:cavity}(b). The background shading illustrates the layer material, with the platinum capping layers forming the cavity, and the thin layer of  \isotope[57]{SS} placed at the guided mode maximum. }
\end{figure}

The calculated probe field intensity profile for the cavity structure Pt \SI{2.8}{\nano\metre}/C \SI{22}{\nano\metre}/\isotope[57]{SS} \SI{0.6}{\nano\metre}/ \mbox{C \SI{22.5}{\nano\metre}}/Pt \SI{15}{\nano\metre} and incidence angle $\theta=\SI{3.35}{\milli\radian}$ [corresponding to the third reflection minimum in Fig.~\ref{fig:cavity}(b)] is presented in Fig.~\ref{fig:field}. This example was chosen such that the thin layer of resonant nuclei is placed at the guided mode maximum. The low-$Z$ layer is thereby used as an inert filler which allows the precise positioning of the resonant layer at the desired depth in the cavity. The guided mode ensures the uniform illumination of all M\"ossbauer nuclei in the layer.

\begin{figure}
\centering
\begin{tikzpicture}[
  level/.style={thick},
  trans/.style={thick, <->},
  darrow/.style={thick,->, dashed}
]
\path (0cm,0cm) rectangle (8.6cm, -4cm);
\coordinate (lh) at (0mm, 5mm);
\coordinate (lw) at (10mm, 0mm);
\coordinate (dh) at (0mm, 0.2mm);
\coordinate (dw) at (1.0mm, 0mm);
\coordinate (bw) at (40mm, 0mm);
\coordinate (bh) at (0mm, 40mm);

\coordinate (i0) at (0mm, 0mm);
\coordinate (i1) at ($(i0) - (bh)$);

\coordinate (j0) at ($(i0) + 0.5*(bw) - 0.2*(lh)$);

\coordinate (m0) at ($(i0) + (bw) + 1.3*(lh)$);
\coordinate (m1) at ($ (m0) - (lh) $);
\coordinate (m2) at ($ (m1) - (lh) $);
\coordinate (m3) at ($ (m2) - (lh) $);

\coordinate (x1) at ($(i1) + 0.05*(bw) + 3*(lh)$);

\coordinate (m4) at ($(i1) + (bw) + 0.5*(lh)$);
\coordinate (m5) at ($(m4) - (lh)$);

\draw[level] (i0) node[left] {$I_e=3/2$} -- ($(i0) + (lw)$);
\draw[darrow] ($(i0) + (lw) + (dw)$) -- node[above] {$\delta$} ($(j0) - (dw)$);
\draw[level] (i1) node[left] {$I_g=1/2$} -- ($(i1) + (lw)$);
\draw[trans] ($(i0) + 0.5*(lw) - (dh)$) -- ($(i1) + 0.5*(lw) + (dh)$);
\node[right] (n1) at ($0.5*(i0) + 0.5*(i1) + 5*(dw)$) {\SI{14.4}{\kilo\electronvolt}};
\draw[level] (j0) -- ($(j0) + (lw)$);
\draw[darrow] ($(j0) + (lw) + (dw)$) -- node[above] {$B_{\mathrm{hf}}$} ($0.5*(m1) + 0.5*(m2)$);


\draw[level] (m0) -- ($(m0) + (lw)$) node[right] {$m_e=3/2$};
\draw[level] (m1) -- ($(m1) + (lw)$) node[right] {$m_e=1/2$};
\draw[level] (m2) -- ($(m2) + (lw)$) node[right] {$m_e=-1/2$};
\draw[level] (m3) -- ($(m3) + (lw)$) node[right] {$m_e=-3/2$};


\draw[level] (m4) -- ($(m4) + (lw)$) node[right] {$m_g=-1/2$};
\draw[level] (m5) -- ($(m5) + (lw)$) node[right] {$m_g=1/2$};

\draw[darrow] ($(i1) + (lw) + 2*(dw)$) -- node[above] {$B_{\mathrm{hf}}$} ($(i1) - 2*(dw) + (bw)$);
\draw[trans] ($(m0) + 0.0*(dw)$) -- ($(m5) + 0.0*(dw)$);
\draw[trans] ($(m1) + 1.2*(dw)$) -- ($(m5) + 1.2*(dw)$);
\draw[trans] ($(m2) + 2.4*(dw)$) -- ($(m5) + 2.4*(dw)$);

\draw[trans] ($(m1) + 6.0*(dw)$) -- ($(m4) + 6.0*(dw)$);
\draw[trans] ($(m2) + 7.2*(dw)$) -- ($(m4) + 7.2*(dw)$);
\draw[trans] ($(m3) + 8.4*(dw)$) -- ($(m4) + 8.4*(dw)$);

\end{tikzpicture}
\caption{\label{fig:lev-scheme}The level scheme of \isotope[57]{Fe}.  The isomer shift $\delta$ shifts all excited states equally, and is due to a monopole interaction with the local electric field. The addition of an external magnetic field $B_{\mathrm{hf}}$ results in a hyperfine splitting of the magnetic sublevels according to their spin projection $m_g$ or $m_e$ on the quantization axis. The six $M1$ transitions are illustrated.}
\end{figure}
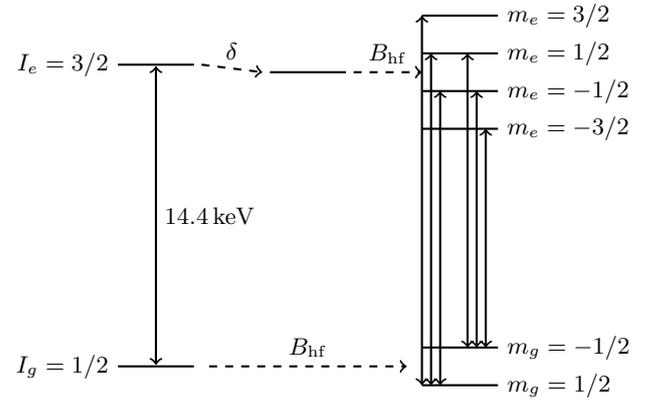

In most cases, the driven M{\" o}ssbauer transitions present hyperfine splittings. Expressed in a multipole expansion, the most significant are the isomer shift, corresponding to the monopole interaction with local electric field, the magnetic splitting due to a dipole interaction with the local magnetic field, and a quadrupole splitting resulting from the interaction with local electric field gradients.  The splittings are determined by the respective magnetic fields or electric field gradients, which are in turn highly sensitive to the local electronic configuration. As most samples are polycrystalline, the hyperfine splittings will be inhomogeneous across the entire sample, but homogeneous within an individual crystal domain.
For the case of the 14.413 keV M\"ossbauer transition in \isotope[57]{Fe}, the natural width is approximately \SI{4.6}{\nano\electronvolt}  and the transition has predominant magnetic dipole character. Magnetic splittings typically  range in the interval $(1 - 50)\gamma_0$, while for example in iron carbides the isomer shift and quadrupole splitting range from $(2-10)\gamma_0$ and $(0-0.55)\gamma_0$~\cite{liu_mossbauer_2016}, respectively, and vary for different crystal structures of the same chemical composition. An illustration of the isomer shift and magnetic splitting for \isotope[57]{Fe} is presented in Fig.~\ref{fig:lev-scheme}. 

The M\"ossbauer transition connects the  \isotope[57]{Fe} ground state with spin $I_g=1/2$ with the first excited state with spin $I_e=3/2$. The magnetic energy shift of each sublevel is given by the expression $m_{e(g)}\mu_{e(g)}B_{\mathrm {hf}}$, where $m_{e(g)}$ are the nuclear spin projections on the quantization axis, $\mu_{e(g)}$ the magnetic moment of the nuclear excited (ground) states, and $B_{\mathrm {hf}}$ the hyperfine magnetic field, respectively. For \isotope[57]{Fe}, the excited (ground) state magnetic splittings account to 3.26~neV/T (5.71~neV/T).

Note that we have used electric dipole transitions in our general derivation, while the transitions in \isotope[57]{Fe} are magnetic dipole. This can be dealt with trivially by making a duality transformation~\cite{buhmannDispersionForces2012}, replacing the incident electric field $\vb{E}$ with the magnetic field $\vb{B}$, electric dipole moment $\wp \vb{d}$ with magnetic moment $\bm{\mu}$, and so forth. In particular the electric dyadic Green's function is replaced with its magnetic dual,
\begin{equation}
  \dyarr{G}(r,r') \to \dyarr{G}^{\star}(r,r').
\end{equation}
In the effective refractive index model of x-ray scattering in matter, the propagation is described via a frequency dependent index of refraction. Outside of resonant interactions, magnetic scattering is orders of magnitude weaker than electronic scattering~\cite{Jackson}. Thus, the magnetic permeability of the layers can be taken to be that of the vacuum, $\mu_0$. 

Therefore, in the notation of Buhmann~\cite{buhmannDispersionForces2012}, we can obtain the dual Green's tensor as
\begin{equation}
  \dyarr{G}^{\star}(r,r') = \frac{1}{\mu_0^2}\overrightarrow{\nabla} \times \dyarr{G}(r,r')\cross \overleftarrow{\nabla}' - \frac{1}{\mu_0}\dyarr{\delta}(r-r').
\end{equation}
The dyadic delta term in the above transformation will modify each individual particle's Lamb shift and line-width equally, and as such can be absorbed into the definitions of $\omega_0$ and $\gamma_0$.

At grazing incidence, the partial Fourier transformed electric Green's function was approximately given by
\begin{equation}
  \dyarr{G}(k_0, \omega_0) \approx \dyarr{\mathds{1}}_\perp \mathcal{G}(k_0, \omega_0),
\end{equation}
i.e. the longitudinal polarization can be neglected. The same will apply to the magnetic component, and therefore, we can take the magnetic dual of this to be
\begin{equation}
  \dyarr{G}^{\star}(k_0, \omega_0) \approx \dyarr{\mathds{1}}_\perp \frac{k_0^2}{\mu_0^2}\mathcal{G}(k_0, \omega_0).
\end{equation}
As seen in Eq.~\eqref{eq:j-gamma-from-g}, the coherent and incoherent collective coupling strength $J,\, \Gamma$ are obtainable directly from the Fourier transformed Green's function. In addition, the quantum optical approach of Heeg and Evers~\cite{heegXrayQuantumOptics2013} demonstrates that the these can also be obtained in terms of the cavity detuning $\Delta_C$ and loss $\kappa$ of the cavity mode excited by the probe field, such that
\begin{equation}
  J+i \Gamma \propto \frac{\Delta_C + i\kappa}{\Delta_C^2 + \kappa^2}.
\end{equation}
In this model, the cavity detuning is minimized when the probe field is incident along one of the reflectivity minima of the cavity shown in Fig.~\ref{fig:cavity}(b), and increases when the angle is shifted away from the minimum. As such, in grazing incidence cavities, the coherent coupling constant $J$ is experimentally controllable by setting the angle of incidence of the probe field.

\subsection{Semi-classical versus quantum models for x-ray quantum optics with M\"ossbauer nuclei \label{sec:history}}
Before presenting some numerical examples, it is instructive to place the present formalism in the context of existing semi-classical and quantum models used in x-ray quantum optics. 
Previous approaches such as by Hannon and Trammel~\cite{hannonMossbauerDiffractionII1969,hannonMossbauerDiffractionIII1974,hannonMossbauerDiffractionQuantum1968,hannonGrazingincidenceAntireflectionFilms1985a,hannonGrazingincidenceAntireflectionFilms1985}, as well as Sturhahn~\cite{sturhahnNuclearResonantSpectroscopy2004,sturhahnTheoreticalAspectsIncoherent1999} have modelled grazing incidence x-ray reflection using a diagrammatic expansion for the photonic scattering. The nuclear interaction is treated semi-classically, with the nuclear transitions taken to be linear dipole oscillators. In this approach, the results of Eq.~\eqref{eq:collective-response} are implicitly modelled, but not explicitly obtainable. Specifically, the response of the nuclei is modelled according to Eq.~\eqref{eq:single-particle-response}, and the re-scattering is implicitly included in the layer matrix formalism for the photonic propagation. However, this approach obscures the collective dynamics of the nuclei. In contrast, the present formalism makes the collective nuclear dynamics features explicit, and allows for the both the coherent and incoherent effects to be investigated separately.

A more recent model, developed by Heeg and Evers~\cite{heegXrayQuantumOptics2013, heegCollectiveEffectsMultiple2015} has focused on the quantum optical perspective, with the emphasis being on the resonant interaction of the nuclei with the cavity mode. A Green's function approach for the scattering part of the Hamiltonian has been developed by Lentrodt \textit{et al.}~\cite{lentrodtInitioQuantumModels2020} and Kong \textit{et al.}~\cite{kongGreenSfunctionFormalism2020}. These works use analytic expressions for the Green's functions in layered media, developed by Toma{\v s}~\cite{tomasGreenFunctionMultilayers1995a} (and later on also by  Johansson \cite{Johansson2011}), which can be expressed in terms of the layer matrix formalism for planar scattering.
Refs.~\cite{lentrodtInitioQuantumModels2020,kongGreenSfunctionFormalism2020} provide a connection between the linear response of the quantum optical model \cite{heegXrayQuantumOptics2013, heegCollectiveEffectsMultiple2015} to the scattering model of Hannon, Trammel, Sturhahn \textit{et al}.  Our work extends this quantum optical Green's function approach  to include inhomogeneous hyperfine parameters,  and provides an insight into how the inhomogeneity will enter into the non-linear dynamics of the system. In the following we provide a few illustrative examples.

\subsection{Gaussian broadening for two level systems}\label{sec:example1}
As an illustrative example, we consider a layer of resonant M\"ossbauer nuclei with a Gaussian distribution of the isomer shift $\delta$. Such Gaussian distributions of hyperfine parameters are typical of amorphous solids~\cite{sharonMagnetismAmorphousFePdP1972,vandiepenMossbauerEffectMagnetic1976,fereyCaracterisationVarieteAmorphie1979}. For simplicity, we will consider the other hyperfine splittings to be negligible.

The isomer shift affects all excited states equally, and the distribution does not affect the dipole vectors or the natural linewidths. As we are taking the other hyperfine splittings to be negligible, all states with a given nuclear spin $I$ are degenerate, and we can model the system as having a single transition. Thus, we can treat the problem as scalar.

In the continuum limit, the coherent average becomes an integral over the distribution of $\delta$,
\begin{equation}
  \sum_n p_n  \to \int_{-\infty}^\infty \dd{\delta} p(\delta),
\end{equation}
where $p(\delta)$ is the probability distribution finding a given value of $\delta$ in the ensemble. The response function $\mathcal{F}(\omega)$ is then given by
\begin{equation}
  \mathcal{F}(\omega) = \int\dd{\delta}p(\delta) \frac{\gamma_0}{\omega - \delta +  i\gamma_0}.
\end{equation}
With a Gaussian distribution, 
\begin{equation}
 p(\delta;\bar{\delta}, \sigma) = \frac{1}{\sqrt{2\pi \sigma^2}}\exp({-\frac{{(\delta-\bar{\delta})}^2}{2\sigma^2}}),
\end{equation}
this evaluates to a Voigt profile~\cite[Eq. 7.19]{NIST:DLMF},
\begin{equation}\label{eq:voigt}
\mathcal{F}(\omega) =\frac{\gamma_0}{\sqrt{2}\sigma}w\left(\frac{\omega - \bar{\delta}+i\gamma_0}{\sqrt{2}\sigma}\right),
\end{equation}
where 
\begin{equation}\label{eq:faddeeva}
  w\left(z\right) = -i\sqrt{\pi}\exp(-z^2)(\erfi(z)-i).
\end{equation}

\begin{figure}
\includegraphics[width=8.6cm]{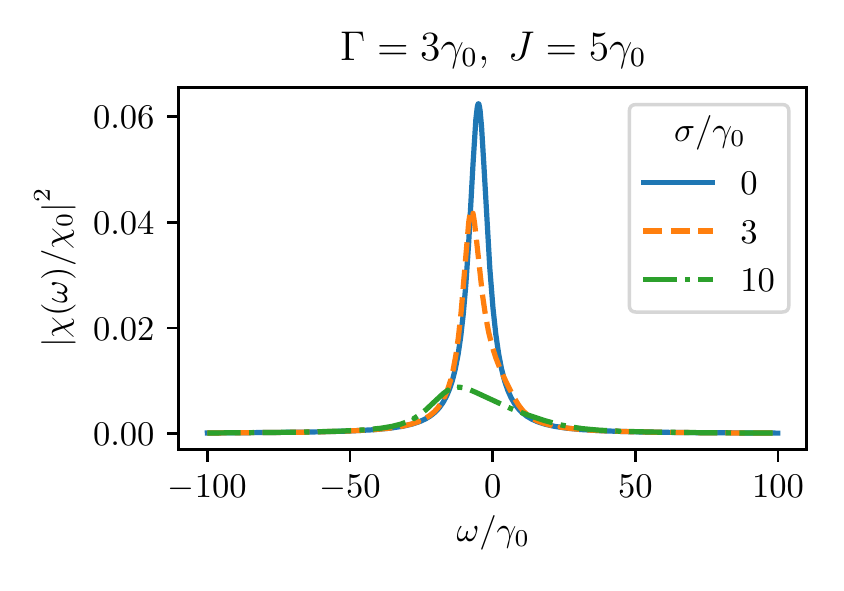}
\caption{\label{fig:single} Collective spectrum $|\chi(\omega)/\chi_0|^2$ as a function of the frequency $\omega$ for $J=5\gamma_0$ and three different distribution widths $\sigma$.   The peak is shifted further as the broadening is increased, and the shape is distorted.}
\end{figure}

As the dipole vectors are all along a single direction, we need only consider the component of susceptibility in this direction, given by
\begin{equation}
\chi(\omega) = \chi_0 \frac{\mathcal{F}(\omega)}{1 + \gamma_0^{-1}(J+i\Gamma) \mathcal{F}(\omega)}.
\end{equation}
The interaction of the coherent collective coupling $J$, and the broadening $\sigma$ is particularly interesting. Unlike the case of purely collective broadening $\Gamma$, for significant distribution widths $\sigma$ the coherent coupling factor $J$ no longer acts as a simple Lamb shift. Indeed, if $J$ were to act as a Lamb shift, from~\eqref{eq:wrong-lamb-shift} one would expect instead a susceptibility of 
\begin{equation}
	\chi(\omega) = \chi_0 F(\omega + J + i \Gamma),
\end{equation}
with a simple translation and broadening. Instead, the peak of the spectrum is shifted slightly further than it would be in the absence of the Gaussian broadening, and is asymmetrically distorted. This is shown in  Fig.~\ref{fig:single} which presents the ratio $|\chi(\omega)/\chi_0|^2$ as a function of $\omega$ for three distributions widths $\sigma$.  

For illustrative purposes, we have used values of $J, \Gamma$ in the range $(0-10)\gamma_0, (3-5)\gamma_0$, typical of x-ray cavities. For example, using \textit{pynuss} to simulate the single line spectrum, we find the cavity in Fig~\ref{fig:cavity} has $J=8.5\gamma_0$ and $\Gamma=3.36\gamma_0$ at an incident angle of \SI{2.32}{\milli\radian}, corresponding to just below the first reflection minimum. Going exactly to the first minimum gives $J=5.5\gamma_0$, and increases $\Gamma$ significantly to $\Gamma=18.6\gamma_0$. At an angle of \SI{3.35}{\milli\radian}, corresponding to the third reflection minimum, we have $J=1.79\gamma_0$ and $\Gamma=3.37\gamma_0$. Typical hyperfine distribution widths are between $(0-5)\gamma_0$. In optical contexts (for example Doppler broadening in atomic clouds, or size inhomogeneity in quantum dots), we would expect that the distribution widths could be substantially larger.

\begin{figure}
\includegraphics[width=8.6cm]{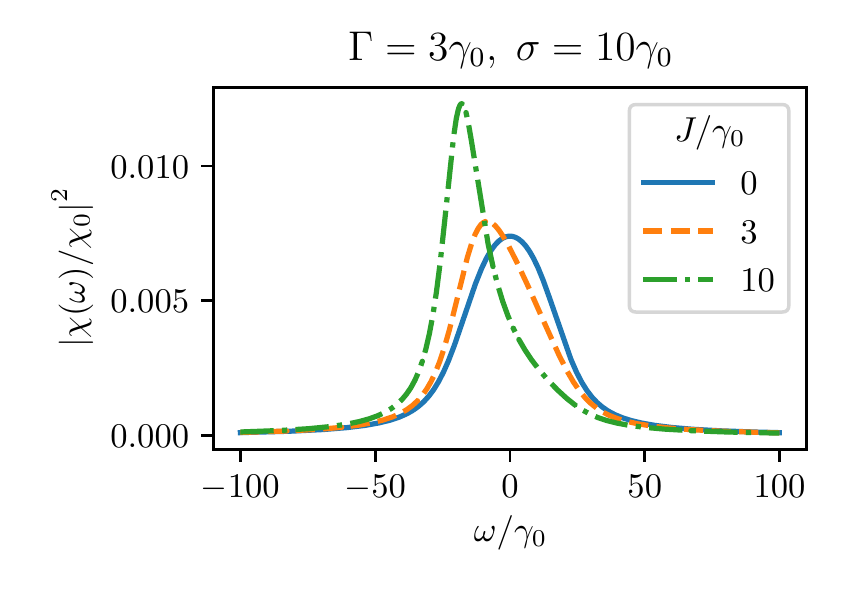}
\caption{\label{fig:single-lamb} Collective spectrum $|\chi(\omega)/\chi_0|^2$ as a function of the frequency $\omega$ for $\sigma=10\gamma_0$ and three different collective coupling values  $J$. With increasing coherent collective coupling $J$ the shape of the spectrum is distorted asymmetrically, and the effective linewidth tends to $\Gamma$.}
\end{figure}

For significant distribution widths $\sigma \gg \gamma_0$, the line shape is that of a broad, almost Gaussian profile. However, as the collective coupling $J$ is increased, as well as being shifted and skewed, the effective linewidth tends to that of the incoherent coupling $\Gamma$.  Figure~\ref{fig:single-lamb} illustrates this behaviour for increasing coherent collective coupling $J$.

\subsection{Interference effects for magnetic splitting}\label{sec:example2}
Let us now consider the case of  magnetic splitting in \isotope[57]{Fe} with no isomer shift and an x-ray field which drives the two $m_e - m_g= 0$ transitions as shown in Fig.~\ref{fig:lev-scheme}. Compared to the  two-level system, the energies of these two transitions are now detuned by $\pm \phi=\pm \frac{1}{2}(\mu_e-\mu_g)B_{\mathrm{hf}}$. Our model Hamiltonian reads
\begin{equation}
  H = \hbar \phi(b_1^\dagger b_1 - b_2^\dagger b_2) 
  -\hbar J(b_1^\dagger + b_2^\dagger)(b_1 + b_2) + \hbar \Omega (b_1^\dagger + b_2^\dagger) + \text{h.c}
\end{equation}
with Lindblad operator
\begin{equation}
  L[\rho] =  -\gamma \sum_{i=1,2}\mathcal{L}[\rho, b_i^\dagger, b_i]- \Gamma\sum_{i,j=1,2}\mathcal{L}[\rho, b_i^\dagger, b_j].
\end{equation}

The superradiant response of such a system was investigated by Kong and P{\' a}lffy~\cite{kongCollectiveRadiationSpectrum2017} using an eigenvalue method. It was found that  if the splitting $\phi$ is less than the incoherent part of the collective coupling $\Gamma$, the contributions from the two transitions interfere. The resulting spectrum has an interference dip in the peak, similar to electromagnetically induced transparency (EIT)~\cite{EITReview2005}, with the collective coupling $J$ playing the role of a control field. In addition, the coherent part of the collective coupling, $J$, was found not to act as a simple Lamb shift, but  in fact non-trivially couple with the magnetic splitting, producing asymmetric, Fano-like spectra.

\begin{figure}[ht]
  \includegraphics[width=8.6 cm]{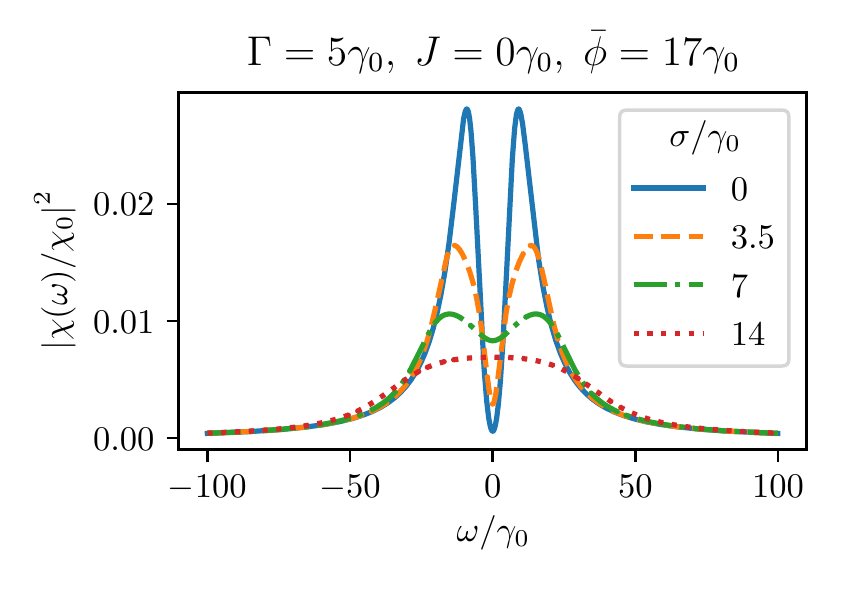}
  \caption{\label{fig:plot_all} Comparison of collective spectrum $|\chi(\omega)/\chi_0|^2$ as a function of the frequency $\omega$ for $\Gamma=5\gamma_0$, $J=0$, with mean splitting $\bar{\phi}=17\gamma_0$ and various values of distribution width $\sigma$. As the distribution width increases, the dip is washed out to a very flat and broad peak.}
\end{figure}

We now consider this system with the addition of a Gaussian distribution of magnetic field strengths across the sites. For a given site with splitting $\phi$, the response matrix is given by~\cite{kongCollectiveRadiationSpectrum2017}
\begin{equation}
\mathcal{F(\omega;\phi)} = \frac{2\gamma_0(\omega+i\gamma_0)}{{(\omega+i\gamma_0)}^2- \phi^2}.
\end{equation}
In the case of a completely uniform magnetic field, the collective susceptibility is therefore given by
\begin{equation}
\chi(\omega) = \chi_0 \frac{2\gamma_0 (\omega+i\gamma_0)}{{(\omega+i\gamma_0)}^2+2(J+i\Gamma)(\omega+i\gamma_0)- \phi^2}.
\end{equation}
This has two poles in the denominator, 
\begin{equation}
\omega_\pm = -i\gamma_0 -J-i\Gamma\pm \sqrt{{(J+i\Gamma)}^2 + \phi^2}.
\end{equation}
When the collective coupling is completely incoherent, $J=0$, the discriminant becomes $\sqrt{\phi^2 - \Gamma^2}$. We can see that if $\phi < \Gamma$, the argument of the square root becomes negative, and the poles become purely imaginary, describing overlapping Lorentzians with differing linewidths. This results in an EIT like dip. This behaviour is illustrated in Fig.~\ref{fig:plot_all} which presents the susceptibility ratio  $|\chi(\omega)/\chi_0|^2$ as a function of $\omega$ for four different values of the Gaussian distribution width $\sigma$. 

If we now consider the magnetic splitting to have a Gaussian distribution of width $\sigma$, and mean $\bar{\phi}$, applying Eqs.~(\ref{eq:single-particle-response}) and~(\ref{eq:voigt}) gives
\begin{equation}
\begin{aligned}
\mathcal{F}(\omega) &= \int \dd{\phi} p(\phi;\bar{\phi},\sigma) F(\omega;\phi) 
\\
&= \frac{\gamma_0}{\sqrt{2}\sigma}\left(w\left(\frac{\omega-\bar{\phi} +i \gamma_0}{\sqrt{2}\sigma}\right) + w\left(\frac{\omega + \bar{\phi} + i\gamma_0}{\sqrt{2}\sigma}\right)\right),
\end{aligned}
\end{equation}
with $w(z)$ given by Equation~\eqref{eq:faddeeva}. The susceptibility is as before given by
\begin{equation}
  \chi(\omega) = \chi_0 \frac{\mathcal{F}(\omega)}{1+ \gamma_0^{-1}(J+i\Gamma)\mathcal{F}(\omega)}.
\end{equation}
The overall envelope of the spectrum resembles that of the homogeneous case, and if the distribution width $\sigma$ is narrow compared with $\Gamma,\bar{\phi}$, we can see that the intensity minimum is still resolvable. However, increasing the distribution width gradually flattens the dip, and results in a flat, broad peak as shown in Fig.~\ref{fig:plot_all}.

\begin{figure}[ht]
  \includegraphics[width=8.6 cm]{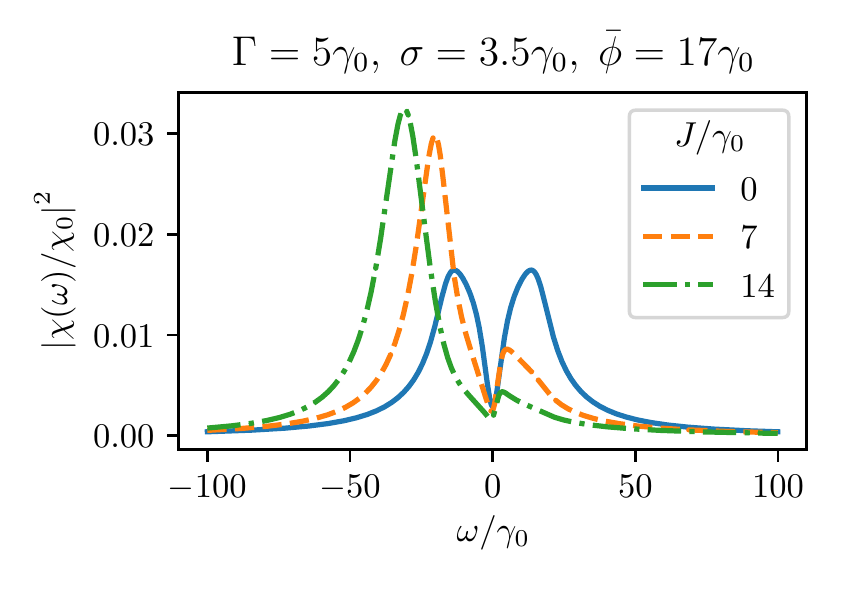}
  \caption{\label{fig:plot_J}Comparison of collective spectrum $|\chi(\omega)/\chi_0|^2$ as a function of the frequency $\omega$ for small incoherent coupling $\Gamma=5\gamma_0$, a small distribution width value $\sigma=3.5\gamma_0$,  mean splitting $\bar{\phi}=17\gamma_0$ and varying values of coherent collective coupling $J$. }
\end{figure}

More interesting is the effect of different strengths of the coherent collective coupling $J$. Rather than acting as a simple Lamb shift, the overall spectral shape is changed. One peak is flattened as the other increases, with large $J$ resulting in a completely asymmetric picture with only a single one of the contributions being resolved. This can be seen in  Fig.~\ref{fig:plot_J} which presents the same dependence as Fig.~\ref{fig:plot_all} but this time for different values of $J$. While the peak locations are somewhat shifted, the shapes are distorted as well, and the location of the minimum is unchanged. This is in contrast to the single line case, where $J$ acts as a pure Lamb shift.

\begin{figure}[ht]
  \includegraphics[width=8.6 cm]{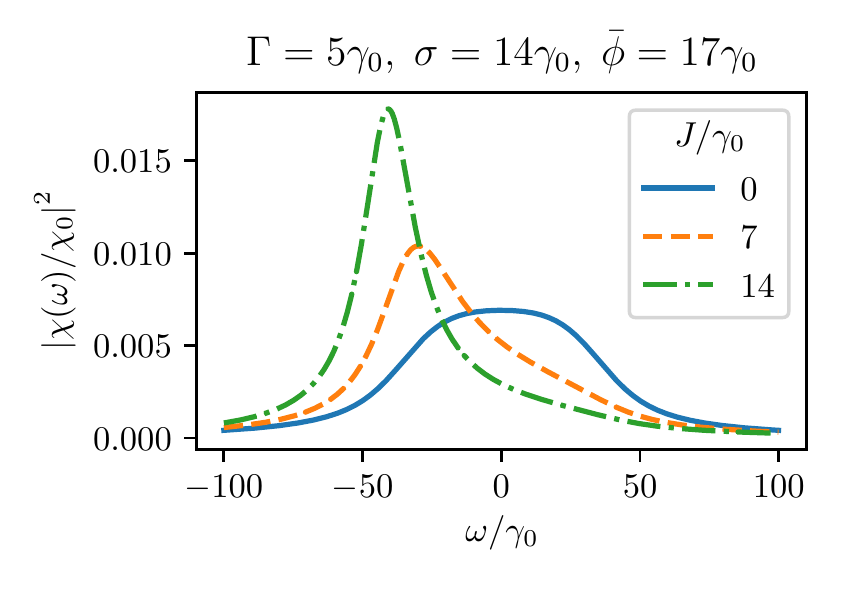}
  \caption{\label{fig:plot_J2} The same as Fig.~\ref{fig:plot_J} but here for a wider distribution width $\sigma=14\gamma_0$. 
  }
\end{figure}

This holds even when the distribution width is large enough that the minimum is not resolved, as shown in Fig.~\ref{fig:plot_J2}. For vanishing collective coupling, $J=0$, the two peaks are merged, and the effective linewidth is very broad. Increasing $J$ results in the left peak growing while the right peak diminishes, and for significant $J$ only the left peak is individually resolved, with the linewidth approaching $2\Gamma$. The result is an increase in the effective resolution of the spectrum, with an energy shift. As the coherent coupling strength is controlled via the angle of incidence of the driving field~\cite{heegXrayQuantumOptics2013,lentrodtInitioQuantumModels2020}, this provides a mechanism for mechanical control of the linewidth of such a sample.

To understand this, we consider the matrix form of the corresponding equation of motion,
\begin{equation}
M\begin{pmatrix}
b_1(\omega) \\ b_2(\omega)
\end{pmatrix} = \begin{pmatrix}
\Omega(\omega) \\ \Omega(\omega)
\end{pmatrix}
\end{equation}
with
\begin{equation}
	M = (J + i \Gamma)\begin{pmatrix}
	1  & 1 
	\\
	1 & 1
	\end{pmatrix}
	+ 
	\begin{pmatrix}
	\phi + i\gamma_0 & 0 
	\\
	0 & -\phi + i \gamma_0
	\end{pmatrix}.
\end{equation}
and $\Omega(\omega) =\hbar^{-1}\wp E_{\text{p}}(\omega)$.

If $J$ is large enough compared with $\phi, \gamma_0$, we may treat the second term as a small perturbation of the first. The eigenvectors of $M$ are then given by
\begin{equation}
\hat{e}_{\pm} = \frac{1}{\sqrt{2}}
\begin{pmatrix}
	1 \\ \pm 1
\end{pmatrix}
+ \mathcal{O}(\phi),
\end{equation}
with eigenvalues
\begin{equation}
\lambda_{+} =2(J + i \Gamma) + \mathcal{O}(\phi),
\quad
\lambda_{-} = \mathcal{O}(\phi)
\end{equation}
The driving term couples to $b_1, b_2$ equally, and is thus proportional to $\hat{e}_+$. Therefore, only the symmetric state $\hat{e}_+$ is strongly driven, and we will expect to see a single peak, with a Lamb shift of $2J$ and a broadening of $2\Gamma$. If the collective broadening $\Gamma$ is significantly lower than the distribution width $\sigma$, we will then see a reduction in the effective linewidth. This has potential applications in samples with significant magnetic texture, with the beam angle of incidence on the sample being used to control the collective coupling, and hence the effective linewidth.

\section{Conclusion \label{sec:concl}}

In this paper, we have examined an extension of the Dicke model for inhomogeneous atoms. We found a compact formula for the susceptibility, in terms of the coherently averaged nuclear/atomic responses, and the collective coupling constants $J,\ \Gamma$. In addition to a collective Lamb shift and broadening, we found that the collective coupling also provides additional cross-couplings between transitions, as well as spontaneously generated coherences. 

Previous work by Kong and P{\'a}lffy~\cite{kongCollectiveRadiationSpectrum2017} has shown that for homogeneous multi-level atoms, the collective coupling does not act as an overall broadening and Lamb shift, and that in particular the coherent coupling $J$ distorts the shape of the line asymmetrically. We have shown that this conclusion holds in the case of inhomogeneous ensembles, and that in addition, the coherent coupling $J$ can counteract the inhomogeneous broadening.

Our work is applicable to one dimensional scattering geometries. For two and three dimensional geometries, if the approximation can be made that the scattering is elastic, with low recoil, then the system can be modelled as quasi one dimensional, and our approach holds. This is naturally the case for M{\" o}ssbauer transitions. However, if this does not hold, the scattered radiation will be emitted in arbitrary directions, and will therefore couple differently to the various transitions depending on the direction of travel. The problem then becomes highly geometrically dependent~\cite{ressayreQuantumTheory1977,svidzinskyCooperativeSpontaneousEmission2008,svidzinskyCooperativeSpontaneousEmission2010}.

In our model, we have only considered linear dynamics. The non-linear dynamics of permutationally invariant systems have been well studied~\cite{shammahOpenQuantumSystems2018,Gegg2016,geggIdenticalEmittersCollective2017}, and can be applied numerically to the sub-ensembles. The Hilbert space of each sub-ensemble scales with $N^{M+1}$ where $M$ is the number of transitions, and $N$ is the maximum excitation number being modelled~\cite{Gegg2016,geggIdenticalEmittersCollective2017}. If the distribution of hyperfine parameters can be modelled in a piecewise constant fashion with $P$ pieces, then the Hilbert space can be modelled as the tensor product of $P$ sub-ensembles, for a final dimension of $N^{P(M+1)}$. The scaling remains polynomial in excitation number, but is exponential in sub-ensemble number, which poses a considerable challenge for numerical simulation.

Finally, we have made the assumption that there is no momentum redistribution during the scattering process. At larger saturations this will not be the case, and the permutational symmetry of the model will be broken. At this point it is not clear whether non-linearities due to the local emission, or momentum redistribution will be more significant. This is the subject of further work.

\begin{acknowledgments}
 This work is part of and supported by the DFG
Collaborative Research Center ``SFB 1225 (ISOQUANT)''. AP acknowledges support from the Heisenberg Program of the DFG. 

\end{acknowledgments}

\bibliography{greens}

\begin{thebibliography}{76}%
\makeatletter
\providecommand \@ifxundefined [1]{%
 \@ifx{#1\undefined}
}%
\providecommand \@ifnum [1]{%
 \ifnum #1\expandafter \@firstoftwo
 \else \expandafter \@secondoftwo
 \fi
}%
\providecommand \@ifx [1]{%
 \ifx #1\expandafter \@firstoftwo
 \else \expandafter \@secondoftwo
 \fi
}%
\providecommand \natexlab [1]{#1}%
\providecommand \enquote  [1]{``#1''}%
\providecommand \bibnamefont  [1]{#1}%
\providecommand \bibfnamefont [1]{#1}%
\providecommand \citenamefont [1]{#1}%
\providecommand \href@noop [0]{\@secondoftwo}%
\providecommand \href [0]{\begingroup \@sanitize@url \@href}%
\providecommand \@href[1]{\@@startlink{#1}\@@href}%
\providecommand \@@href[1]{\endgroup#1\@@endlink}%
\providecommand \@sanitize@url [0]{\catcode `\\12\catcode `\$12\catcode
  `\&12\catcode `\#12\catcode `\^12\catcode `\_12\catcode `\%12\relax}%
\providecommand \@@startlink[1]{}%
\providecommand \@@endlink[0]{}%
\providecommand \url  [0]{\begingroup\@sanitize@url \@url }%
\providecommand \@url [1]{\endgroup\@href {#1}{\urlprefix }}%
\providecommand \urlprefix  [0]{URL }%
\providecommand \Eprint [0]{\href }%
\providecommand \doibase [0]{https://doi.org/}%
\providecommand \selectlanguage [0]{\@gobble}%
\providecommand \bibinfo  [0]{\@secondoftwo}%
\providecommand \bibfield  [0]{\@secondoftwo}%
\providecommand \translation [1]{[#1]}%
\providecommand \BibitemOpen [0]{}%
\providecommand \bibitemStop [0]{}%
\providecommand \bibitemNoStop [0]{.\EOS\space}%
\providecommand \EOS [0]{\spacefactor3000\relax}%
\providecommand \BibitemShut  [1]{\csname bibitem#1\endcsname}%
\let\auto@bib@innerbib\@empty
\bibitem [{\citenamefont
  {Dicke}(1954)}]{dickeCoherenceSpontaneousRadiation1954}%
  \BibitemOpen
  \bibfield  {author} {\bibinfo {author} {\bibfnamefont {R.~H.}\ \bibnamefont
  {Dicke}},\ }\bibfield  {title} {\bibinfo {title} {Coherence in {{Spontaneous
  Radiation Processes}}},\ }\href {https://doi.org/10.1103/PhysRev.93.99}
  {\bibfield  {journal} {\bibinfo  {journal} {Physical Review}\ }\textbf
  {\bibinfo {volume} {93}},\ \bibinfo {pages} {99} (\bibinfo {year}
  {1954})}\BibitemShut {NoStop}%
\bibitem [{\citenamefont {Rehler}\ and\ \citenamefont
  {Eberly}(1971)}]{Rehler1971}%
  \BibitemOpen
  \bibfield  {author} {\bibinfo {author} {\bibfnamefont {N.~E.}\ \bibnamefont
  {Rehler}}\ and\ \bibinfo {author} {\bibfnamefont {J.~H.}\ \bibnamefont
  {Eberly}},\ }\bibfield  {title} {\bibinfo {title} {Superradiance},\ }\href
  {https://doi.org/10.1103/PhysRevA.3.1735} {\bibfield  {journal} {\bibinfo
  {journal} {Phys. Rev. A}\ }\textbf {\bibinfo {volume} {3}},\ \bibinfo {pages}
  {1735} (\bibinfo {year} {1971})}\BibitemShut {NoStop}%
\bibitem [{\citenamefont {Gross}\ and\ \citenamefont
  {Haroche}(1982)}]{Haroche1982}%
  \BibitemOpen
  \bibfield  {author} {\bibinfo {author} {\bibfnamefont {M.}~\bibnamefont
  {Gross}}\ and\ \bibinfo {author} {\bibfnamefont {S.}~\bibnamefont
  {Haroche}},\ }\bibfield  {title} {\bibinfo {title} {Superradiance: An essay
  on the theory of collective spontaneous emission},\ }\href
  {https://doi.org/https://doi.org/10.1016/0370-1573(82)90102-8} {\bibfield
  {journal} {\bibinfo  {journal} {Physics Reports}\ }\textbf {\bibinfo {volume}
  {93}},\ \bibinfo {pages} {301 } (\bibinfo {year} {1982})}\BibitemShut
  {NoStop}%
\bibitem [{\citenamefont {Leonardi}\ \emph {et~al.}(1986)\citenamefont
  {Leonardi}, \citenamefont {Persico},\ and\ \citenamefont
  {Vetri}}]{Leonardi1986review}%
  \BibitemOpen
  \bibfield  {author} {\bibinfo {author} {\bibfnamefont {C.}~\bibnamefont
  {Leonardi}}, \bibinfo {author} {\bibfnamefont {F.}~\bibnamefont {Persico}},\
  and\ \bibinfo {author} {\bibfnamefont {G.}~\bibnamefont {Vetri}},\
  }\href@noop {} {\bibfield  {journal} {\bibinfo  {journal} {Riv. Nuovo
  Cimento}\ }\textbf {\bibinfo {volume} {9}},\ \bibinfo {pages} {1 (issue 4)}
  (\bibinfo {year} {1986})}\BibitemShut {NoStop}%
\bibitem [{\citenamefont {Skribanowitz}\ \emph {et~al.}(1973)\citenamefont
  {Skribanowitz}, \citenamefont {Herman}, \citenamefont {MacGillivray},\ and\
  \citenamefont {Feld}}]{Skribanowitz1973}%
  \BibitemOpen
  \bibfield  {author} {\bibinfo {author} {\bibfnamefont {N.}~\bibnamefont
  {Skribanowitz}}, \bibinfo {author} {\bibfnamefont {I.~P.}\ \bibnamefont
  {Herman}}, \bibinfo {author} {\bibfnamefont {J.~C.}\ \bibnamefont
  {MacGillivray}},\ and\ \bibinfo {author} {\bibfnamefont {M.~S.}\ \bibnamefont
  {Feld}},\ }\bibfield  {title} {\bibinfo {title} {Observation of dicke
  superradiance in optically pumped hf gas},\ }\href
  {https://doi.org/10.1103/PhysRevLett.30.309} {\bibfield  {journal} {\bibinfo
  {journal} {Phys. Rev. Lett.}\ }\textbf {\bibinfo {volume} {30}},\ \bibinfo
  {pages} {309} (\bibinfo {year} {1973})}\BibitemShut {NoStop}%
\bibitem [{\citenamefont {Inouye}\ \emph {et~al.}(1999)\citenamefont {Inouye},
  \citenamefont {Chikkatur}, \citenamefont {Stamper-Kurn}, \citenamefont
  {Stenger}, \citenamefont {Pritchard},\ and\ \citenamefont
  {Ketterle}}]{Inouye1999}%
  \BibitemOpen
  \bibfield  {author} {\bibinfo {author} {\bibfnamefont {S.}~\bibnamefont
  {Inouye}}, \bibinfo {author} {\bibfnamefont {A.~P.}\ \bibnamefont
  {Chikkatur}}, \bibinfo {author} {\bibfnamefont {D.~M.}\ \bibnamefont
  {Stamper-Kurn}}, \bibinfo {author} {\bibfnamefont {J.}~\bibnamefont
  {Stenger}}, \bibinfo {author} {\bibfnamefont {D.~E.}\ \bibnamefont
  {Pritchard}},\ and\ \bibinfo {author} {\bibfnamefont {W.}~\bibnamefont
  {Ketterle}},\ }\bibfield  {title} {\bibinfo {title} {Superradiant rayleigh
  scattering from a bose-einstein condensate},\ }\href
  {https://doi.org/10.1126/science.285.5427.571} {\bibfield  {journal}
  {\bibinfo  {journal} {Science}\ }\textbf {\bibinfo {volume} {285}},\ \bibinfo
  {pages} {571} (\bibinfo {year} {1999})}\BibitemShut {NoStop}%
\bibitem [{\citenamefont {Baumann}\ \emph {et~al.}(2010)\citenamefont
  {Baumann}, \citenamefont {Guerlin}, \citenamefont {Brennecke},\ and\
  \citenamefont {Esslinger}}]{Baumann2010}%
  \BibitemOpen
  \bibfield  {author} {\bibinfo {author} {\bibfnamefont {K.}~\bibnamefont
  {Baumann}}, \bibinfo {author} {\bibfnamefont {C.}~\bibnamefont {Guerlin}},
  \bibinfo {author} {\bibfnamefont {F.}~\bibnamefont {Brennecke}},\ and\
  \bibinfo {author} {\bibfnamefont {T.}~\bibnamefont {Esslinger}},\ }\href@noop
  {} {\bibfield  {journal} {\bibinfo  {journal} {Nature}\ }\textbf {\bibinfo
  {volume} {464}},\ \bibinfo {pages} {1301} (\bibinfo {year}
  {2010})}\BibitemShut {NoStop}%
\bibitem [{\citenamefont {Scheibner}\ \emph {et~al.}(2007)\citenamefont
  {Scheibner}, \citenamefont {Schmidt}, \citenamefont {Worschech},
  \citenamefont {Forchel}, \citenamefont {Bacher}, \citenamefont {Passow},\
  and\ \citenamefont {Hommel}}]{Scheibner2007}%
  \BibitemOpen
  \bibfield  {author} {\bibinfo {author} {\bibfnamefont {M.}~\bibnamefont
  {Scheibner}}, \bibinfo {author} {\bibfnamefont {T.}~\bibnamefont {Schmidt}},
  \bibinfo {author} {\bibfnamefont {L.}~\bibnamefont {Worschech}}, \bibinfo
  {author} {\bibfnamefont {A.}~\bibnamefont {Forchel}}, \bibinfo {author}
  {\bibfnamefont {G.}~\bibnamefont {Bacher}}, \bibinfo {author} {\bibfnamefont
  {T.}~\bibnamefont {Passow}},\ and\ \bibinfo {author} {\bibfnamefont
  {D.}~\bibnamefont {Hommel}},\ }\bibfield  {title} {\bibinfo {title}
  {Superradiance of quantum dots},\ }\href {https://doi.org/10.1038/nphys494}
  {\bibfield  {journal} {\bibinfo  {journal} {Nature Physics}\ }\textbf
  {\bibinfo {volume} {3}},\ \bibinfo {pages} {106} (\bibinfo {year}
  {2007})}\BibitemShut {NoStop}%
\bibitem [{\citenamefont {Angerer}\ \emph {et~al.}(2018)\citenamefont
  {Angerer}, \citenamefont {Streltsov}, \citenamefont {Astner}, \citenamefont
  {Putz}, \citenamefont {Sumiya}, \citenamefont {Onoda}, \citenamefont {Isoya},
  \citenamefont {Munro}, \citenamefont {Nemoto}, \citenamefont {Schiedmeyer},\
  and\ \citenamefont {Majer}}]{Angerer2018}%
  \BibitemOpen
  \bibfield  {author} {\bibinfo {author} {\bibfnamefont {A.}~\bibnamefont
  {Angerer}}, \bibinfo {author} {\bibfnamefont {K.}~\bibnamefont {Streltsov}},
  \bibinfo {author} {\bibfnamefont {T.}~\bibnamefont {Astner}}, \bibinfo
  {author} {\bibfnamefont {S.}~\bibnamefont {Putz}}, \bibinfo {author}
  {\bibfnamefont {H.}~\bibnamefont {Sumiya}}, \bibinfo {author} {\bibfnamefont
  {S.}~\bibnamefont {Onoda}}, \bibinfo {author} {\bibfnamefont
  {J.}~\bibnamefont {Isoya}}, \bibinfo {author} {\bibfnamefont {W.~J.}\
  \bibnamefont {Munro}}, \bibinfo {author} {\bibfnamefont {K.}~\bibnamefont
  {Nemoto}}, \bibinfo {author} {\bibfnamefont {K.}~\bibnamefont
  {Schiedmeyer}},\ and\ \bibinfo {author} {\bibfnamefont {J.}~\bibnamefont
  {Majer}},\ }\href@noop {} {\bibfield  {journal} {\bibinfo  {journal} {Nature
  Phys.}\ }\textbf {\bibinfo {volume} {14}},\ \bibinfo {pages} {1168} (\bibinfo
  {year} {2018})}\BibitemShut {NoStop}%
\bibitem [{\citenamefont {Ara{\'u}jo}\ \emph {et~al.}(2016)\citenamefont
  {Ara{\'u}jo}, \citenamefont {Kre{\v s}i{\'c}}, \citenamefont {Kaiser},\ and\
  \citenamefont {Guerin}}]{araujoSuperradianceLargeDilute2016a}%
  \BibitemOpen
  \bibfield  {author} {\bibinfo {author} {\bibfnamefont {M.~O.}\ \bibnamefont
  {Ara{\'u}jo}}, \bibinfo {author} {\bibfnamefont {I.}~\bibnamefont {Kre{\v
  s}i{\'c}}}, \bibinfo {author} {\bibfnamefont {R.}~\bibnamefont {Kaiser}},\
  and\ \bibinfo {author} {\bibfnamefont {W.}~\bibnamefont {Guerin}},\
  }\bibfield  {title} {\bibinfo {title} {Superradiance in a {{Large}} and
  {{Dilute Cloud}} of {{Cold Atoms}} in the {{Linear}}-{{Optics Regime}}},\
  }\href {https://doi.org/10.1103/PhysRevLett.117.073002} {\bibfield  {journal}
  {\bibinfo  {journal} {Phys. Rev. Lett.}\ }\textbf {\bibinfo {volume} {117}},\
  \bibinfo {pages} {073002} (\bibinfo {year} {2016})}\BibitemShut {NoStop}%
\bibitem [{\citenamefont {Cottier}\ \emph {et~al.}(2018)\citenamefont
  {Cottier}, \citenamefont {Kaiser},\ and\ \citenamefont
  {Bachelard}}]{cottierRoleDisorderSuper2018a}%
  \BibitemOpen
  \bibfield  {author} {\bibinfo {author} {\bibfnamefont {F.}~\bibnamefont
  {Cottier}}, \bibinfo {author} {\bibfnamefont {R.}~\bibnamefont {Kaiser}},\
  and\ \bibinfo {author} {\bibfnamefont {R.}~\bibnamefont {Bachelard}},\
  }\bibfield  {title} {\bibinfo {title} {Role of disorder in super- and
  subradiance of cold atomic clouds},\ }\href
  {https://doi.org/10.1103/PhysRevA.98.013622} {\bibfield  {journal} {\bibinfo
  {journal} {Phys. Rev. A}\ }\textbf {\bibinfo {volume} {98}},\ \bibinfo
  {pages} {013622} (\bibinfo {year} {2018})}\BibitemShut {NoStop}%
\bibitem [{\citenamefont {Adams}\ \emph {et~al.}(2013)\citenamefont {Adams},
  \citenamefont {Buth}, \citenamefont {Cavaletto}, \citenamefont {Evers},
  \citenamefont {Harman}, \citenamefont {Keitel}, \citenamefont {P{\'a}lffy},
  \citenamefont {Pic{\'o}n}, \citenamefont {R{\"o}hlsberger}, \citenamefont
  {Rostovtsev},\ and\ \citenamefont {Tamasaku}}]{xrayReview2013}%
  \BibitemOpen
  \bibfield  {author} {\bibinfo {author} {\bibfnamefont {B.~W.}\ \bibnamefont
  {Adams}}, \bibinfo {author} {\bibfnamefont {C.}~\bibnamefont {Buth}},
  \bibinfo {author} {\bibfnamefont {S.~M.}\ \bibnamefont {Cavaletto}}, \bibinfo
  {author} {\bibfnamefont {J.}~\bibnamefont {Evers}}, \bibinfo {author}
  {\bibfnamefont {Z.}~\bibnamefont {Harman}}, \bibinfo {author} {\bibfnamefont
  {C.~H.}\ \bibnamefont {Keitel}}, \bibinfo {author} {\bibfnamefont
  {A.}~\bibnamefont {P{\'a}lffy}}, \bibinfo {author} {\bibfnamefont
  {A.}~\bibnamefont {Pic{\'o}n}}, \bibinfo {author} {\bibfnamefont
  {R.}~\bibnamefont {R{\"o}hlsberger}}, \bibinfo {author} {\bibfnamefont
  {Y.}~\bibnamefont {Rostovtsev}},\ and\ \bibinfo {author} {\bibfnamefont
  {K.}~\bibnamefont {Tamasaku}},\ }\href@noop {} {\bibfield  {journal}
  {\bibinfo  {journal} {J. Mod. Opt.}\ }\textbf {\bibinfo {volume} {60}},\
  \bibinfo {pages} {2} (\bibinfo {year} {2013})}\BibitemShut {NoStop}%
\bibitem [{\citenamefont {R{\"o}hlsberger}\ \emph {et~al.}(2010)\citenamefont
  {R{\"o}hlsberger}, \citenamefont {Schlage}, \citenamefont {Sahoo},
  \citenamefont {Couet},\ and\ \citenamefont
  {R{\"u}ffer}}]{rohlsbergerCollectiveLambShift2010}%
  \BibitemOpen
  \bibfield  {author} {\bibinfo {author} {\bibfnamefont {R.}~\bibnamefont
  {R{\"o}hlsberger}}, \bibinfo {author} {\bibfnamefont {K.}~\bibnamefont
  {Schlage}}, \bibinfo {author} {\bibfnamefont {B.}~\bibnamefont {Sahoo}},
  \bibinfo {author} {\bibfnamefont {S.}~\bibnamefont {Couet}},\ and\ \bibinfo
  {author} {\bibfnamefont {R.}~\bibnamefont {R{\"u}ffer}},\ }\bibfield  {title}
  {\bibinfo {title} {Collective lamb shift in single-photon superradiance},\
  }\href {https://doi.org/10.1126/science.1187770} {\bibfield  {journal}
  {\bibinfo  {journal} {Science}\ }\textbf {\bibinfo {volume} {328}},\ \bibinfo
  {pages} {1248} (\bibinfo {year} {2010})}\BibitemShut {NoStop}%
\bibitem [{\citenamefont {Chumakov}\ \emph {et~al.}(2017)\citenamefont
  {Chumakov}, \citenamefont {Baron}, \citenamefont {Sergueev}, \citenamefont
  {Strohm}, \citenamefont {Leupold}, \citenamefont {Shvyd'ko}, \citenamefont
  {Smirnov}, \citenamefont {R{\"u}ffer}, \citenamefont {Inubushi},
  \citenamefont {Yabashi}, \citenamefont {Tono}, \citenamefont {Kudo},\ and\
  \citenamefont {Ishikawa}}]{Chumakov2017}%
  \BibitemOpen
  \bibfield  {author} {\bibinfo {author} {\bibfnamefont {A.~I.}\ \bibnamefont
  {Chumakov}}, \bibinfo {author} {\bibfnamefont {A.~Q.~R.}\ \bibnamefont
  {Baron}}, \bibinfo {author} {\bibfnamefont {I.}~\bibnamefont {Sergueev}},
  \bibinfo {author} {\bibfnamefont {C.}~\bibnamefont {Strohm}}, \bibinfo
  {author} {\bibfnamefont {O.}~\bibnamefont {Leupold}}, \bibinfo {author}
  {\bibfnamefont {Y.}~\bibnamefont {Shvyd'ko}}, \bibinfo {author}
  {\bibfnamefont {G.~V.}\ \bibnamefont {Smirnov}}, \bibinfo {author}
  {\bibfnamefont {R.}~\bibnamefont {R{\"u}ffer}}, \bibinfo {author}
  {\bibfnamefont {Y.}~\bibnamefont {Inubushi}}, \bibinfo {author}
  {\bibfnamefont {M.}~\bibnamefont {Yabashi}}, \bibinfo {author} {\bibfnamefont
  {K.}~\bibnamefont {Tono}}, \bibinfo {author} {\bibfnamefont {T.}~\bibnamefont
  {Kudo}},\ and\ \bibinfo {author} {\bibfnamefont {T.}~\bibnamefont
  {Ishikawa}},\ }\bibfield  {title} {\bibinfo {title} {Superradiance of an
  ensemble of nuclei excited by a free electron laser},\ }\href
  {https://doi.org/10.1038/s41567-017-0001-z} {\bibfield  {journal} {\bibinfo
  {journal} {Nature Physics}\ }\textbf {\bibinfo {volume} {14}},\ \bibinfo
  {pages} {261} (\bibinfo {year} {2017})}\BibitemShut {NoStop}%
\bibitem [{\citenamefont {Reimann}\ \emph {et~al.}(2015)\citenamefont
  {Reimann}, \citenamefont {Alt}, \citenamefont {Kampschulte}, \citenamefont
  {Macha}, \citenamefont {Ratschbacher}, \citenamefont {Thau}, \citenamefont
  {Yoon},\ and\ \citenamefont {Meschede}}]{Reimann2015}%
  \BibitemOpen
  \bibfield  {author} {\bibinfo {author} {\bibfnamefont {R.}~\bibnamefont
  {Reimann}}, \bibinfo {author} {\bibfnamefont {W.}~\bibnamefont {Alt}},
  \bibinfo {author} {\bibfnamefont {T.}~\bibnamefont {Kampschulte}}, \bibinfo
  {author} {\bibfnamefont {T.}~\bibnamefont {Macha}}, \bibinfo {author}
  {\bibfnamefont {L.}~\bibnamefont {Ratschbacher}}, \bibinfo {author}
  {\bibfnamefont {N.}~\bibnamefont {Thau}}, \bibinfo {author} {\bibfnamefont
  {S.}~\bibnamefont {Yoon}},\ and\ \bibinfo {author} {\bibfnamefont
  {D.}~\bibnamefont {Meschede}},\ }\bibfield  {title} {\bibinfo {title}
  {Cavity-modified collective rayleigh scattering of two atoms},\ }\href
  {https://doi.org/10.1103/PhysRevLett.114.023601} {\bibfield  {journal}
  {\bibinfo  {journal} {Phys. Rev. Lett.}\ }\textbf {\bibinfo {volume} {114}},\
  \bibinfo {pages} {023601} (\bibinfo {year} {2015})}\BibitemShut {NoStop}%
\bibitem [{\citenamefont {Kim}\ \emph {et~al.}(2018)\citenamefont {Kim},
  \citenamefont {Yang}, \citenamefont {Oh},\ and\ \citenamefont
  {An}}]{Kim2018}%
  \BibitemOpen
  \bibfield  {author} {\bibinfo {author} {\bibfnamefont {J.}~\bibnamefont
  {Kim}}, \bibinfo {author} {\bibfnamefont {D.}~\bibnamefont {Yang}}, \bibinfo
  {author} {\bibfnamefont {S.-h.}\ \bibnamefont {Oh}},\ and\ \bibinfo {author}
  {\bibfnamefont {K.}~\bibnamefont {An}},\ }\bibfield  {title} {\bibinfo
  {title} {Coherent single-atom superradiance},\ }\href
  {https://doi.org/10.1126/science.aar2179} {\bibfield  {journal} {\bibinfo
  {journal} {Science}\ }\textbf {\bibinfo {volume} {359}},\ \bibinfo {pages}
  {662} (\bibinfo {year} {2018})}\BibitemShut {NoStop}%
\bibitem [{\citenamefont {Scully}\ \emph {et~al.}(2006)\citenamefont {Scully},
  \citenamefont {Fry}, \citenamefont {Ooi},\ and\ \citenamefont
  {W{\'o}dkiewicz}}]{scullyDirectedSpontaneousEmission2006}%
  \BibitemOpen
  \bibfield  {author} {\bibinfo {author} {\bibfnamefont {M.~O.}\ \bibnamefont
  {Scully}}, \bibinfo {author} {\bibfnamefont {E.~S.}\ \bibnamefont {Fry}},
  \bibinfo {author} {\bibfnamefont {C.~H.~R.}\ \bibnamefont {Ooi}},\ and\
  \bibinfo {author} {\bibfnamefont {K.}~\bibnamefont {W{\'o}dkiewicz}},\
  }\bibfield  {title} {\bibinfo {title} {Directed {{Spontaneous Emission}} from
  an {{Extended Ensemble}} of \${{N}}\$ {{Atoms}}: {{Timing Is Everything}}},\
  }\href {https://doi.org/10.1103/PhysRevLett.96.010501} {\bibfield  {journal}
  {\bibinfo  {journal} {Phys. Rev. Lett.}\ }\textbf {\bibinfo {volume} {96}},\
  \bibinfo {pages} {010501} (\bibinfo {year} {2006})}\BibitemShut {NoStop}%
\bibitem [{\citenamefont {{de Oliveira}}\ \emph {et~al.}(2014)\citenamefont
  {{de Oliveira}}, \citenamefont {Mendes}, \citenamefont {Martins},
  \citenamefont {Saldanha}, \citenamefont {Tabosa},\ and\ \citenamefont
  {Felinto}}]{deoliveiraSinglephotonSuperradianceCold2014}%
  \BibitemOpen
  \bibfield  {author} {\bibinfo {author} {\bibfnamefont {R.~A.}\ \bibnamefont
  {{de Oliveira}}}, \bibinfo {author} {\bibfnamefont {M.~S.}\ \bibnamefont
  {Mendes}}, \bibinfo {author} {\bibfnamefont {W.~S.}\ \bibnamefont {Martins}},
  \bibinfo {author} {\bibfnamefont {P.~L.}\ \bibnamefont {Saldanha}}, \bibinfo
  {author} {\bibfnamefont {J.~W.~R.}\ \bibnamefont {Tabosa}},\ and\ \bibinfo
  {author} {\bibfnamefont {D.}~\bibnamefont {Felinto}},\ }\bibfield  {title}
  {\bibinfo {title} {Single-photon superradiance in cold atoms},\ }\href
  {https://doi.org/10.1103/PhysRevA.90.023848} {\bibfield  {journal} {\bibinfo
  {journal} {Phys. Rev. A}\ }\textbf {\bibinfo {volume} {90}},\ \bibinfo
  {pages} {023848} (\bibinfo {year} {2014})}\BibitemShut {NoStop}%
\bibitem [{\citenamefont {Rajabi}\ and\ \citenamefont
  {Houde}(2016)}]{Rajabi2016}%
  \BibitemOpen
  \bibfield  {author} {\bibinfo {author} {\bibfnamefont {F.}~\bibnamefont
  {Rajabi}}\ and\ \bibinfo {author} {\bibfnamefont {M.}~\bibnamefont {Houde}},\
  }\bibfield  {title} {\bibinfo {title} {Dicke's superradaince in astrophysics.
  i. the 21 cm line},\ }\href {https://doi.org/10.3847/0004-637x/826/2/216}
  {\bibfield  {journal} {\bibinfo  {journal} {The Astrophysical Journal}\
  }\textbf {\bibinfo {volume} {826}},\ \bibinfo {pages} {216} (\bibinfo {year}
  {2016})}\BibitemShut {NoStop}%
\bibitem [{\citenamefont {Vieira}\ \emph {et~al.}(2021)\citenamefont {Vieira},
  \citenamefont {Pardal}, \citenamefont {Mendon\'ca},\ and\ \citenamefont
  {Fonseca}}]{Vieira2021}%
  \BibitemOpen
  \bibfield  {author} {\bibinfo {author} {\bibfnamefont {C.}~\bibnamefont
  {Vieira}}, \bibinfo {author} {\bibfnamefont {M.}~\bibnamefont {Pardal}},
  \bibinfo {author} {\bibfnamefont {J.~T.}\ \bibnamefont {Mendon\'ca}},\ and\
  \bibinfo {author} {\bibfnamefont {A.~R.}\ \bibnamefont {Fonseca}},\
  }\href@noop {} {\bibfield  {journal} {\bibinfo  {journal} {Nature Phys.}\
  }\textbf {\bibinfo {volume} {17}},\ \bibinfo {pages} {99} (\bibinfo {year}
  {2021})}\BibitemShut {NoStop}%
\bibitem [{\citenamefont {Arecchi}\ \emph {et~al.}(1973)\citenamefont
  {Arecchi}, \citenamefont {Gilmore},\ and\ \citenamefont
  {Kim}}]{arecchiCoherentStatesRlevel1973}%
  \BibitemOpen
  \bibfield  {author} {\bibinfo {author} {\bibfnamefont {P.~T.}\ \bibnamefont
  {Arecchi}}, \bibinfo {author} {\bibfnamefont {E.}~\bibnamefont {Gilmore}},\
  and\ \bibinfo {author} {\bibfnamefont {D.~M.}\ \bibnamefont {Kim}},\
  }\bibfield  {title} {\bibinfo {title} {Coherent states for r-level atoms},\
  }\href {https://doi.org/10.1007/BF02827264} {\bibfield  {journal} {\bibinfo
  {journal} {Lettere al Nuovo Cimento}\ }\textbf {\bibinfo {volume} {6}},\
  \bibinfo {pages} {219} (\bibinfo {year} {1973})}\BibitemShut {NoStop}%
\bibitem [{\citenamefont {M{\"u}ller}\ \emph {et~al.}(2005)\citenamefont
  {M{\"u}ller}, \citenamefont {Miniatura}, \citenamefont {Wilkowski},
  \citenamefont {Kaiser},\ and\ \citenamefont
  {Delande}}]{mullerMultipleScatteringPhotons2005a}%
  \BibitemOpen
  \bibfield  {author} {\bibinfo {author} {\bibfnamefont {C.~A.}\ \bibnamefont
  {M{\"u}ller}}, \bibinfo {author} {\bibfnamefont {C.}~\bibnamefont
  {Miniatura}}, \bibinfo {author} {\bibfnamefont {D.}~\bibnamefont
  {Wilkowski}}, \bibinfo {author} {\bibfnamefont {R.}~\bibnamefont {Kaiser}},\
  and\ \bibinfo {author} {\bibfnamefont {D.}~\bibnamefont {Delande}},\
  }\bibfield  {title} {\bibinfo {title} {Multiple scattering of photons by
  atomic hyperfine multiplets},\ }\href
  {https://doi.org/10.1103/PhysRevA.72.053405} {\bibfield  {journal} {\bibinfo
  {journal} {Phys. Rev. A}\ }\textbf {\bibinfo {volume} {72}},\ \bibinfo
  {pages} {053405} (\bibinfo {year} {2005})}\BibitemShut {NoStop}%
\bibitem [{\citenamefont {Gegg}\ and\ \citenamefont
  {Richter}(2016)}]{Gegg2016}%
  \BibitemOpen
  \bibfield  {author} {\bibinfo {author} {\bibfnamefont {M.}~\bibnamefont
  {Gegg}}\ and\ \bibinfo {author} {\bibfnamefont {M.}~\bibnamefont {Richter}},\
  }\bibfield  {title} {\bibinfo {title} {Efficient and exact numerical approach
  for many multi-level systems in open system {CQED}},\ }\href
  {https://doi.org/10.1088/1367-2630/18/4/043037} {\bibfield  {journal}
  {\bibinfo  {journal} {New Journal of Physics}\ }\textbf {\bibinfo {volume}
  {18}},\ \bibinfo {pages} {043037} (\bibinfo {year} {2016})}\BibitemShut
  {NoStop}%
\bibitem [{\citenamefont {Sutherland}\ and\ \citenamefont
  {Robicheaux}(2017)}]{sutherlandSuperradianceInvertedMultilevel2017}%
  \BibitemOpen
  \bibfield  {author} {\bibinfo {author} {\bibfnamefont {R.~T.}\ \bibnamefont
  {Sutherland}}\ and\ \bibinfo {author} {\bibfnamefont {F.}~\bibnamefont
  {Robicheaux}},\ }\bibfield  {title} {\bibinfo {title} {Superradiance in
  inverted multilevel atomic clouds},\ }\href
  {https://doi.org/10.1103/PhysRevA.95.033839} {\bibfield  {journal} {\bibinfo
  {journal} {Physical Review A}\ }\textbf {\bibinfo {volume} {95}},\  (\bibinfo
  {year} {2017})}\BibitemShut {NoStop}%
\bibitem [{\citenamefont {Gegg}(2017)}]{geggIdenticalEmittersCollective2017}%
  \BibitemOpen
  \bibfield  {author} {\bibinfo {author} {\bibfnamefont {M.}~\bibnamefont
  {Gegg}},\ }\emph {\bibinfo {title} {Identical {{Emitters}}, {{Collective
  Effects}} and {{Dissipation}} in {{Quantum Optics}}}},\ \href@noop {} {Ph.D.
  thesis},\ \bibinfo  {school} {Technische Universit{\"a}t Berlin} (\bibinfo
  {year} {2017})\BibitemShut {NoStop}%
\bibitem [{\citenamefont {Kong}\ and\ \citenamefont
  {P{\'a}lffy}(2017)}]{kongCollectiveRadiationSpectrum2017}%
  \BibitemOpen
  \bibfield  {author} {\bibinfo {author} {\bibfnamefont {X.}~\bibnamefont
  {Kong}}\ and\ \bibinfo {author} {\bibfnamefont {A.}~\bibnamefont
  {P{\'a}lffy}},\ }\bibfield  {title} {\bibinfo {title} {Collective radiation
  spectrum for ensembles with {{Zeeman}} splitting in single-photon
  superradiance},\ }\href {https://doi.org/10.1103/PhysRevA.96.033819}
  {\bibfield  {journal} {\bibinfo  {journal} {Phys. Rev. A}\ }\textbf {\bibinfo
  {volume} {96}},\ \bibinfo {pages} {033819} (\bibinfo {year}
  {2017})}\BibitemShut {NoStop}%
\bibitem [{\citenamefont {Das}\ \emph {et~al.}(2018{\natexlab{a}})\citenamefont
  {Das}, \citenamefont {Elfving}, \citenamefont {Reiter},\ and\ \citenamefont
  {S\o{}rensen}}]{SumantaI-2018}%
  \BibitemOpen
  \bibfield  {author} {\bibinfo {author} {\bibfnamefont {S.}~\bibnamefont
  {Das}}, \bibinfo {author} {\bibfnamefont {V.~E.}\ \bibnamefont {Elfving}},
  \bibinfo {author} {\bibfnamefont {F.}~\bibnamefont {Reiter}},\ and\ \bibinfo
  {author} {\bibfnamefont {A.~S.}\ \bibnamefont {S\o{}rensen}},\ }\bibfield
  {title} {\bibinfo {title} {Photon scattering from a system of multilevel
  quantum emitters. i. formalism},\ }\href
  {https://doi.org/10.1103/PhysRevA.97.043837} {\bibfield  {journal} {\bibinfo
  {journal} {Phys. Rev. A}\ }\textbf {\bibinfo {volume} {97}},\ \bibinfo
  {pages} {043837} (\bibinfo {year} {2018}{\natexlab{a}})}\BibitemShut
  {NoStop}%
\bibitem [{\citenamefont {Agarwal}(1971)}]{Agarwal1971}%
  \BibitemOpen
  \bibfield  {author} {\bibinfo {author} {\bibfnamefont {G.~S.}\ \bibnamefont
  {Agarwal}},\ }\bibfield  {title} {\bibinfo {title} {Master-equation approach
  to spontaneous emission. iii. many-body aspects of emission from two-level
  atoms and the effect of inhomogeneous broadening},\ }\href
  {https://doi.org/10.1103/PhysRevA.4.1791} {\bibfield  {journal} {\bibinfo
  {journal} {Phys. Rev. A}\ }\textbf {\bibinfo {volume} {4}},\ \bibinfo {pages}
  {1791} (\bibinfo {year} {1971})}\BibitemShut {NoStop}%
\bibitem [{\citenamefont {Eberly}(1971{\natexlab{a}})}]{Eberly1971}%
  \BibitemOpen
  \bibfield  {author} {\bibinfo {author} {\bibfnamefont {J.~H.}\ \bibnamefont
  {Eberly}},\ }\href@noop {} {\bibfield  {journal} {\bibinfo  {journal} {Lett.
  Nuovo Cimento}\ }\textbf {\bibinfo {volume} {1}},\ \bibinfo {pages} {182}
  (\bibinfo {year} {1971}{\natexlab{a}})}\BibitemShut {NoStop}%
\bibitem [{\citenamefont {Eberly}(1971{\natexlab{b}})}]{Eberly1971-Acta}%
  \BibitemOpen
  \bibfield  {author} {\bibinfo {author} {\bibfnamefont {J.~H.}\ \bibnamefont
  {Eberly}},\ }\href@noop {} {\bibfield  {journal} {\bibinfo  {journal} {Acta
  Phys. Pol. A}\ }\textbf {\bibinfo {volume} {39}},\ \bibinfo {pages} {633}
  (\bibinfo {year} {1971}{\natexlab{b}})}\BibitemShut {NoStop}%
\bibitem [{\citenamefont {Jodoin}\ and\ \citenamefont
  {Mandel}(1974)}]{Jodoin1974}%
  \BibitemOpen
  \bibfield  {author} {\bibinfo {author} {\bibfnamefont {R.}~\bibnamefont
  {Jodoin}}\ and\ \bibinfo {author} {\bibfnamefont {L.}~\bibnamefont
  {Mandel}},\ }\bibfield  {title} {\bibinfo {title} {Superradiance in an
  inhomogeneously broadened atomic system},\ }\href
  {https://doi.org/10.1103/PhysRevA.9.873} {\bibfield  {journal} {\bibinfo
  {journal} {Phys. Rev. A}\ }\textbf {\bibinfo {volume} {9}},\ \bibinfo {pages}
  {873} (\bibinfo {year} {1974})}\BibitemShut {NoStop}%
\bibitem [{\citenamefont {Leonardi}\ and\ \citenamefont
  {Vaglica}(1982)}]{Leonardi1982}%
  \BibitemOpen
  \bibfield  {author} {\bibinfo {author} {\bibfnamefont {C.}~\bibnamefont
  {Leonardi}}\ and\ \bibinfo {author} {\bibfnamefont {A.}~\bibnamefont
  {Vaglica}},\ }\href@noop {} {\bibfield  {journal} {\bibinfo  {journal} {Il
  Nuovo Cimento B}\ }\textbf {\bibinfo {volume} {67}},\ \bibinfo {pages} {256}
  (\bibinfo {year} {1982})}\BibitemShut {NoStop}%
\bibitem [{\citenamefont {Spano}\ and\ \citenamefont
  {Mukamel}(1989)}]{MolAggr1989}%
  \BibitemOpen
  \bibfield  {author} {\bibinfo {author} {\bibfnamefont {F.~C.}\ \bibnamefont
  {Spano}}\ and\ \bibinfo {author} {\bibfnamefont {S.}~\bibnamefont
  {Mukamel}},\ }\bibfield  {title} {\bibinfo {title} {Superradiance in
  molecular aggregates},\ }\href {https://doi.org/10.1063/1.457174} {\bibfield
  {journal} {\bibinfo  {journal} {The Journal of Chemical Physics}\ }\textbf
  {\bibinfo {volume} {91}},\ \bibinfo {pages} {683} (\bibinfo {year}
  {1989})}\BibitemShut {NoStop}%
\bibitem [{\citenamefont {Temnov}\ and\ \citenamefont
  {Woggon}(2005)}]{Temnov2005}%
  \BibitemOpen
  \bibfield  {author} {\bibinfo {author} {\bibfnamefont {V.~V.}\ \bibnamefont
  {Temnov}}\ and\ \bibinfo {author} {\bibfnamefont {U.}~\bibnamefont
  {Woggon}},\ }\bibfield  {title} {\bibinfo {title} {Superradiance and
  subradiance in an inhomogeneously broadened ensemble of two-level systems
  coupled to a low-$q$ cavity},\ }\href
  {https://doi.org/10.1103/PhysRevLett.95.243602} {\bibfield  {journal}
  {\bibinfo  {journal} {Phys. Rev. Lett.}\ }\textbf {\bibinfo {volume} {95}},\
  \bibinfo {pages} {243602} (\bibinfo {year} {2005})}\BibitemShut {NoStop}%
\bibitem [{\citenamefont {Bonifacio}\ and\ \citenamefont
  {Lugiato}(1975)}]{Bonifacio1975}%
  \BibitemOpen
  \bibfield  {author} {\bibinfo {author} {\bibfnamefont {R.}~\bibnamefont
  {Bonifacio}}\ and\ \bibinfo {author} {\bibfnamefont {L.~A.}\ \bibnamefont
  {Lugiato}},\ }\bibfield  {title} {\bibinfo {title} {Cooperative radiation
  processes in two-level systems: Superfluorescence},\ }\href
  {https://doi.org/10.1103/PhysRevA.11.1507} {\bibfield  {journal} {\bibinfo
  {journal} {Phys. Rev. A}\ }\textbf {\bibinfo {volume} {11}},\ \bibinfo
  {pages} {1507} (\bibinfo {year} {1975})}\BibitemShut {NoStop}%
\bibitem [{\citenamefont {Haake}\ \emph {et~al.}(1980)\citenamefont {Haake},
  \citenamefont {Haus}, \citenamefont {King}, \citenamefont {Schr{\"o}der},\
  and\ \citenamefont {Glauber}}]{Haake1980}%
  \BibitemOpen
  \bibfield  {author} {\bibinfo {author} {\bibfnamefont {F.}~\bibnamefont
  {Haake}}, \bibinfo {author} {\bibfnamefont {J.}~\bibnamefont {Haus}},
  \bibinfo {author} {\bibfnamefont {H.}~\bibnamefont {King}}, \bibinfo {author}
  {\bibfnamefont {G.}~\bibnamefont {Schr{\"o}der}},\ and\ \bibinfo {author}
  {\bibfnamefont {R.}~\bibnamefont {Glauber}},\ }\bibfield  {title} {\bibinfo
  {title} {Delay-time statistics and inhomogeneous line broadening in
  superfluorescence},\ }\href {https://doi.org/10.1103/PhysRevLett.45.558}
  {\bibfield  {journal} {\bibinfo  {journal} {Phys. Rev. Lett.}\ }\textbf
  {\bibinfo {volume} {45}},\ \bibinfo {pages} {558} (\bibinfo {year}
  {1980})}\BibitemShut {NoStop}%
\bibitem [{\citenamefont {Ishikawa}\ \emph {et~al.}(2016)\citenamefont
  {Ishikawa}, \citenamefont {Miyajima}, \citenamefont {Ashida}, \citenamefont
  {Itoh},\ and\ \citenamefont {Ishihara}}]{Ishikawa2016}%
  \BibitemOpen
  \bibfield  {author} {\bibinfo {author} {\bibfnamefont {A.}~\bibnamefont
  {Ishikawa}}, \bibinfo {author} {\bibfnamefont {K.}~\bibnamefont {Miyajima}},
  \bibinfo {author} {\bibfnamefont {M.}~\bibnamefont {Ashida}}, \bibinfo
  {author} {\bibfnamefont {T.}~\bibnamefont {Itoh}},\ and\ \bibinfo {author}
  {\bibfnamefont {H.}~\bibnamefont {Ishihara}},\ }\bibfield  {title} {\bibinfo
  {title} {Theory of superfluorescence in highly inhomogeneous quantum
  systems},\ }\href {https://doi.org/10.7566/JPSJ.85.034703} {\bibfield
  {journal} {\bibinfo  {journal} {Journal of the Physical Society of Japan}\
  }\textbf {\bibinfo {volume} {85}},\ \bibinfo {pages} {034703} (\bibinfo
  {year} {2016})},\ \Eprint
  {https://arxiv.org/abs/https://doi.org/10.7566/JPSJ.85.034703}
  {https://doi.org/10.7566/JPSJ.85.034703} \BibitemShut {NoStop}%
\bibitem [{\citenamefont {Roof}\ \emph {et~al.}(2016)\citenamefont {Roof},
  \citenamefont {Kemp}, \citenamefont {Havey},\ and\ \citenamefont
  {Sokolov}}]{roofObservationSinglePhotonSuperradiance2016}%
  \BibitemOpen
  \bibfield  {author} {\bibinfo {author} {\bibfnamefont {S.~J.}\ \bibnamefont
  {Roof}}, \bibinfo {author} {\bibfnamefont {K.~J.}\ \bibnamefont {Kemp}},
  \bibinfo {author} {\bibfnamefont {M.~D.}\ \bibnamefont {Havey}},\ and\
  \bibinfo {author} {\bibfnamefont {I.~M.}\ \bibnamefont {Sokolov}},\
  }\bibfield  {title} {\bibinfo {title} {Observation of {{Single}}-{{Photon
  Superradiance}} and the {{Cooperative Lamb Shift}} in an {{Extended Sample}}
  of {{Cold Atoms}}},\ }\href {https://doi.org/10.1103/PhysRevLett.117.073003}
  {\bibfield  {journal} {\bibinfo  {journal} {Phys. Rev. Lett.}\ }\textbf
  {\bibinfo {volume} {117}},\ \bibinfo {pages} {073003} (\bibinfo {year}
  {2016})}\BibitemShut {NoStop}%
\bibitem [{\citenamefont {Ruostekoski}\ and\ \citenamefont
  {Javanainen}(2016)}]{ruostekoskiEmergenceCorrelatedOptics2016a}%
  \BibitemOpen
  \bibfield  {author} {\bibinfo {author} {\bibfnamefont {J.}~\bibnamefont
  {Ruostekoski}}\ and\ \bibinfo {author} {\bibfnamefont {J.}~\bibnamefont
  {Javanainen}},\ }\bibfield  {title} {\bibinfo {title} {Emergence of
  correlated optics in one-dimensional waveguides for classical and quantum
  atomic gases},\ }\href {https://doi.org/10.1103/PhysRevLett.117.143602}
  {\bibfield  {journal} {\bibinfo  {journal} {Phys. Rev. Lett.}\ }\textbf
  {\bibinfo {volume} {117}},\ \bibinfo {pages} {143602} (\bibinfo {year}
  {2016})}\BibitemShut {NoStop}%
\bibitem [{\citenamefont {{Asenjo-Garcia}}\ \emph {et~al.}(2017)\citenamefont
  {{Asenjo-Garcia}}, \citenamefont {Hood}, \citenamefont {Chang},\ and\
  \citenamefont
  {Kimble}}]{asenjo-garciaAtomlightInteractionsQuasionedimensional2017}%
  \BibitemOpen
  \bibfield  {author} {\bibinfo {author} {\bibfnamefont {A.}~\bibnamefont
  {{Asenjo-Garcia}}}, \bibinfo {author} {\bibfnamefont {J.~D.}\ \bibnamefont
  {Hood}}, \bibinfo {author} {\bibfnamefont {D.~E.}\ \bibnamefont {Chang}},\
  and\ \bibinfo {author} {\bibfnamefont {H.~J.}\ \bibnamefont {Kimble}},\
  }\bibfield  {title} {\bibinfo {title} {Atom-light interactions in
  quasi-one-dimensional nanostructures: {{A Green}}'s-function perspective},\
  }\href {https://doi.org/10.1103/PhysRevA.95.033818} {\bibfield  {journal}
  {\bibinfo  {journal} {Physical Review A}\ }\textbf {\bibinfo {volume} {95}},\
   (\bibinfo {year} {2017})}\BibitemShut {NoStop}%
\bibitem [{\citenamefont {Das}\ \emph {et~al.}(2018{\natexlab{b}})\citenamefont
  {Das}, \citenamefont {Elfving}, \citenamefont {Reiter},\ and\ \citenamefont
  {S\o{}rensen}}]{SumantaII-2018}%
  \BibitemOpen
  \bibfield  {author} {\bibinfo {author} {\bibfnamefont {S.}~\bibnamefont
  {Das}}, \bibinfo {author} {\bibfnamefont {V.~E.}\ \bibnamefont {Elfving}},
  \bibinfo {author} {\bibfnamefont {F.}~\bibnamefont {Reiter}},\ and\ \bibinfo
  {author} {\bibfnamefont {A.~S.}\ \bibnamefont {S\o{}rensen}},\ }\bibfield
  {title} {\bibinfo {title} {Photon scattering from a system of multilevel
  quantum emitters. ii. application to emitters coupled to a one-dimensional
  waveguide},\ }\href {https://doi.org/10.1103/PhysRevA.97.043838} {\bibfield
  {journal} {\bibinfo  {journal} {Phys. Rev. A}\ }\textbf {\bibinfo {volume}
  {97}},\ \bibinfo {pages} {043838} (\bibinfo {year}
  {2018}{\natexlab{b}})}\BibitemShut {NoStop}%
\bibitem [{\citenamefont {Benedict}\ and\ \citenamefont
  {Trifonov}(1988)}]{benedictCoherentReflectionSuperradiation1988}%
  \BibitemOpen
  \bibfield  {author} {\bibinfo {author} {\bibfnamefont {M.~G.}\ \bibnamefont
  {Benedict}}\ and\ \bibinfo {author} {\bibfnamefont {E.~D.}\ \bibnamefont
  {Trifonov}},\ }\bibfield  {title} {\bibinfo {title} {Coherent reflection as
  superradiation from the boundary of a resonant medium},\ }\href
  {https://doi.org/10.1103/PhysRevA.38.2854} {\bibfield  {journal} {\bibinfo
  {journal} {Phys. Rev. A}\ }\textbf {\bibinfo {volume} {38}},\ \bibinfo
  {pages} {2854} (\bibinfo {year} {1988})}\BibitemShut {NoStop}%
\bibitem [{\citenamefont {Samson}\ \emph {et~al.}(1990)\citenamefont {Samson},
  \citenamefont {Logvin},\ and\ \citenamefont
  {Turovets}}]{samsonInducedSuperradianceThin1990}%
  \BibitemOpen
  \bibfield  {author} {\bibinfo {author} {\bibfnamefont {A.~M.}\ \bibnamefont
  {Samson}}, \bibinfo {author} {\bibfnamefont {Y.~A.}\ \bibnamefont {Logvin}},\
  and\ \bibinfo {author} {\bibfnamefont {S.~I.}\ \bibnamefont {Turovets}},\
  }\bibfield  {title} {\bibinfo {title} {Induced superradiance in a thin film
  of two-level atoms},\ }\href {https://doi.org/10.1016/0030-4018(90)90346-U}
  {\bibfield  {journal} {\bibinfo  {journal} {Optics Communications}\ }\textbf
  {\bibinfo {volume} {78}},\ \bibinfo {pages} {208} (\bibinfo {year}
  {1990})}\BibitemShut {NoStop}%
\bibitem [{\citenamefont {Lim}\ \emph {et~al.}(2004)\citenamefont {Lim},
  \citenamefont {Bjorklund}, \citenamefont {Spano},\ and\ \citenamefont
  {Bardeen}}]{limExcitonDelocalizationSuperradiance2004}%
  \BibitemOpen
  \bibfield  {author} {\bibinfo {author} {\bibfnamefont {S.-H.}\ \bibnamefont
  {Lim}}, \bibinfo {author} {\bibfnamefont {T.~G.}\ \bibnamefont {Bjorklund}},
  \bibinfo {author} {\bibfnamefont {F.~C.}\ \bibnamefont {Spano}},\ and\
  \bibinfo {author} {\bibfnamefont {C.~J.}\ \bibnamefont {Bardeen}},\
  }\bibfield  {title} {\bibinfo {title} {Exciton {{Delocalization}} and
  {{Superradiance}} in {{Tetracene Thin Films}} and {{Nanoaggregates}}},\
  }\href {https://doi.org/10.1103/PhysRevLett.92.107402} {\bibfield  {journal}
  {\bibinfo  {journal} {Phys. Rev. Lett.}\ }\textbf {\bibinfo {volume} {92}},\
  \bibinfo {pages} {107402} (\bibinfo {year} {2004})}\BibitemShut {NoStop}%
\bibitem [{\citenamefont {Heeg}\ and\ \citenamefont
  {Evers}(2015)}]{heegCollectiveEffectsMultiple2015}%
  \BibitemOpen
  \bibfield  {author} {\bibinfo {author} {\bibfnamefont {K.~P.}\ \bibnamefont
  {Heeg}}\ and\ \bibinfo {author} {\bibfnamefont {J.}~\bibnamefont {Evers}},\
  }\bibfield  {title} {\bibinfo {title} {Collective effects between multiple
  nuclear ensembles in an x-ray cavity-{{QED}} setup},\ }\href
  {https://doi.org/10.1103/PhysRevA.91.063803} {\bibfield  {journal} {\bibinfo
  {journal} {Physical Review A}\ }\textbf {\bibinfo {volume} {91}},\  (\bibinfo
  {year} {2015})}\BibitemShut {NoStop}%
\bibitem [{\citenamefont {Lentrodt}\ \emph {et~al.}(2020)\citenamefont
  {Lentrodt}, \citenamefont {Heeg}, \citenamefont {Keitel},\ and\ \citenamefont
  {Evers}}]{lentrodtInitioQuantumModels2020}%
  \BibitemOpen
  \bibfield  {author} {\bibinfo {author} {\bibfnamefont {D.}~\bibnamefont
  {Lentrodt}}, \bibinfo {author} {\bibfnamefont {K.~P.}\ \bibnamefont {Heeg}},
  \bibinfo {author} {\bibfnamefont {C.~H.}\ \bibnamefont {Keitel}},\ and\
  \bibinfo {author} {\bibfnamefont {J.}~\bibnamefont {Evers}},\ }\bibfield
  {title} {\bibinfo {title} {Ab initio quantum models for thin-film x-ray
  cavity {{QED}}},\ }\href {https://doi.org/10.1103/PhysRevResearch.2.023396}
  {\bibfield  {journal} {\bibinfo  {journal} {Phys. Rev. Research}\ }\textbf
  {\bibinfo {volume} {2}},\ \bibinfo {pages} {023396} (\bibinfo {year}
  {2020})}\BibitemShut {NoStop}%
\bibitem [{\citenamefont {Gruner}\ and\ \citenamefont
  {Welsch}(1996)}]{grunerGreenfunctionApproachRadiationfield1996}%
  \BibitemOpen
  \bibfield  {author} {\bibinfo {author} {\bibfnamefont {T.}~\bibnamefont
  {Gruner}}\ and\ \bibinfo {author} {\bibfnamefont {D.-G.}\ \bibnamefont
  {Welsch}},\ }\bibfield  {title} {\bibinfo {title} {Green-function approach to
  the radiation-field quantization for homogeneous and inhomogeneous
  {{Kramers}}-{{Kronig}} dielectrics},\ }\href
  {https://doi.org/10.1103/PhysRevA.53.1818} {\bibfield  {journal} {\bibinfo
  {journal} {Physical Review A}\ }\textbf {\bibinfo {volume} {53}},\ \bibinfo
  {pages} {1818} (\bibinfo {year} {1996})}\BibitemShut {NoStop}%
\bibitem [{\citenamefont {Svidzinsky}\ and\ \citenamefont
  {Chang}(2008)}]{svidzinskyCooperativeSpontaneousEmission2008}%
  \BibitemOpen
  \bibfield  {author} {\bibinfo {author} {\bibfnamefont {A.}~\bibnamefont
  {Svidzinsky}}\ and\ \bibinfo {author} {\bibfnamefont {J.-T.}\ \bibnamefont
  {Chang}},\ }\bibfield  {title} {\bibinfo {title} {Cooperative spontaneous
  emission as a many-body eigenvalue problem},\ }\href
  {https://doi.org/10.1103/PhysRevA.77.043833} {\bibfield  {journal} {\bibinfo
  {journal} {Physical Review A}\ }\textbf {\bibinfo {volume} {77}},\  (\bibinfo
  {year} {2008})}\BibitemShut {NoStop}%
\bibitem [{\citenamefont {Svidzinsky}\ \emph {et~al.}(2010)\citenamefont
  {Svidzinsky}, \citenamefont {Chang},\ and\ \citenamefont
  {Scully}}]{svidzinskyCooperativeSpontaneousEmission2010}%
  \BibitemOpen
  \bibfield  {author} {\bibinfo {author} {\bibfnamefont {A.~A.}\ \bibnamefont
  {Svidzinsky}}, \bibinfo {author} {\bibfnamefont {J.-T.}\ \bibnamefont
  {Chang}},\ and\ \bibinfo {author} {\bibfnamefont {M.~O.}\ \bibnamefont
  {Scully}},\ }\bibfield  {title} {\bibinfo {title} {Cooperative {{Spontaneous
  Emission}} of {{N Atoms}}: {{Many}}-{{Body Eigenstates}}, the {{Effect}} of
  {{Virtual Lamb Shift Processes}}, and {{Analogy}} with {{Radiation}} of {{N
  Classical Oscillators}}},\ }\href
  {https://doi.org/10.1103/PhysRevA.81.053821} {\bibfield  {journal} {\bibinfo
  {journal} {Phys. Rev. A}\ }\textbf {\bibinfo {volume} {81}},\ \bibinfo
  {pages} {053821} (\bibinfo {year} {2010})}\BibitemShut {NoStop}%
\bibitem [{\citenamefont {Kong}\ \emph {et~al.}(2020)\citenamefont {Kong},
  \citenamefont {Chang},\ and\ \citenamefont
  {P{\'a}lffy}}]{kongGreenSfunctionFormalism2020}%
  \BibitemOpen
  \bibfield  {author} {\bibinfo {author} {\bibfnamefont {X.}~\bibnamefont
  {Kong}}, \bibinfo {author} {\bibfnamefont {D.~E.}\ \bibnamefont {Chang}},\
  and\ \bibinfo {author} {\bibfnamefont {A.}~\bibnamefont {P{\'a}lffy}},\
  }\bibfield  {title} {\bibinfo {title} {Green's-function formalism for
  resonant interaction of x rays with nuclei in structured media},\ }\href
  {https://doi.org/10.1103/PhysRevA.102.033710} {\bibfield  {journal} {\bibinfo
   {journal} {Phys. Rev. A}\ }\textbf {\bibinfo {volume} {102}},\ \bibinfo
  {pages} {033710} (\bibinfo {year} {2020})}\BibitemShut {NoStop}%
\bibitem [{\citenamefont {Kurucz}\ and\ \citenamefont
  {M{\o}lmer}(2010)}]{kuruczMultilevelHolsteinPrimakoffApproximation2010}%
  \BibitemOpen
  \bibfield  {author} {\bibinfo {author} {\bibfnamefont {Z.}~\bibnamefont
  {Kurucz}}\ and\ \bibinfo {author} {\bibfnamefont {K.}~\bibnamefont
  {M{\o}lmer}},\ }\bibfield  {title} {\bibinfo {title} {Multilevel
  {{Holstein}}-{{Primakoff}} approximation and its application to atomic spin
  squeezing and ensemble quantum memories},\ }\href
  {https://doi.org/10.1103/PhysRevA.81.032314} {\bibfield  {journal} {\bibinfo
  {journal} {Phys. Rev. A}\ }\textbf {\bibinfo {volume} {81}},\ \bibinfo
  {pages} {032314} (\bibinfo {year} {2010})}\BibitemShut {NoStop}%
\bibitem [{\citenamefont {Shammah}\ \emph {et~al.}(2018)\citenamefont
  {Shammah}, \citenamefont {Ahmed}, \citenamefont {Lambert}, \citenamefont
  {De~Liberato},\ and\ \citenamefont {Nori}}]{shammahOpenQuantumSystems2018}%
  \BibitemOpen
  \bibfield  {author} {\bibinfo {author} {\bibfnamefont {N.}~\bibnamefont
  {Shammah}}, \bibinfo {author} {\bibfnamefont {S.}~\bibnamefont {Ahmed}},
  \bibinfo {author} {\bibfnamefont {N.}~\bibnamefont {Lambert}}, \bibinfo
  {author} {\bibfnamefont {S.}~\bibnamefont {De~Liberato}},\ and\ \bibinfo
  {author} {\bibfnamefont {F.}~\bibnamefont {Nori}},\ }\bibfield  {title}
  {\bibinfo {title} {Open quantum systems with local and collective incoherent
  processes: {{Efficient}} numerical simulation using permutational
  invariance},\ }\href {https://doi.org/10.1103/PhysRevA.98.063815} {\bibfield
  {journal} {\bibinfo  {journal} {Phys. Rev. A}\ }\textbf {\bibinfo {volume}
  {98}},\ \bibinfo {pages} {063815} (\bibinfo {year} {2018})}\BibitemShut
  {NoStop}%
\bibitem [{\citenamefont
  {R{\"o}hlsberger}(2004)}]{rohlsbergerNuclearCondensedMatter2004}%
  \BibitemOpen
  \bibfield  {author} {\bibinfo {author} {\bibfnamefont {R.}~\bibnamefont
  {R{\"o}hlsberger}},\ }\href@noop {} {\emph {\bibinfo {title} {Nuclear
  {{Condensed Matter Physics}} with {{Synchrotron Radiation}}: {{Basic
  Principles}}, {{Methodology}} and {{Applications}}}}}\ (\bibinfo  {publisher}
  {{Springer Science \& Business Media}},\ \bibinfo {year} {2004})\BibitemShut
  {NoStop}%
\bibitem [{\citenamefont {Hannon}\ \emph {et~al.}(1988)\citenamefont {Hannon},
  \citenamefont {Trammell}, \citenamefont {Blume},\ and\ \citenamefont
  {Gibbs}}]{hannonResonantExchange1988}%
  \BibitemOpen
  \bibfield  {author} {\bibinfo {author} {\bibfnamefont {J.~P.}\ \bibnamefont
  {Hannon}}, \bibinfo {author} {\bibfnamefont {G.~T.}\ \bibnamefont
  {Trammell}}, \bibinfo {author} {\bibfnamefont {M.}~\bibnamefont {Blume}},\
  and\ \bibinfo {author} {\bibfnamefont {D.}~\bibnamefont {Gibbs}},\ }\bibfield
   {title} {\bibinfo {title} {x-ray resonance exchange scattering},\ }\href
  {https://doi.org/10.1103/PhysRevLett.61.1245} {\bibfield  {journal} {\bibinfo
   {journal} {Phys. Rev. Lett.}\ }\textbf {\bibinfo {volume} {61}},\ \bibinfo
  {pages} {1245} (\bibinfo {year} {1988})}\BibitemShut {NoStop}%
\bibitem [{\citenamefont {Scully}\ and\ \citenamefont
  {Svidzinsky}(2010)}]{ScullyLambShift2010}%
  \BibitemOpen
  \bibfield  {author} {\bibinfo {author} {\bibfnamefont {M.~O.}\ \bibnamefont
  {Scully}}\ and\ \bibinfo {author} {\bibfnamefont {A.~A.}\ \bibnamefont
  {Svidzinsky}},\ }\bibfield  {title} {\bibinfo {title} {The lamb
  shift{\textemdash}yesterday, today, and tomorrow},\ }\href
  {https://doi.org/10.1126/science.1190737} {\bibfield  {journal} {\bibinfo
  {journal} {Science}\ }\textbf {\bibinfo {volume} {328}},\ \bibinfo {pages}
  {1239} (\bibinfo {year} {2010})}\BibitemShut {NoStop}%
\bibitem [{\citenamefont {Wen}\ \emph {et~al.}(2019)\citenamefont {Wen},
  \citenamefont {Lin}, \citenamefont {Kockum}, \citenamefont {Suri},
  \citenamefont {Ian}, \citenamefont {Chen}, \citenamefont {Mao}, \citenamefont
  {Chiu}, \citenamefont {Delsing}, \citenamefont {Nori}, \citenamefont {Lin},\
  and\ \citenamefont {Hoi}}]{wen2019}%
  \BibitemOpen
  \bibfield  {author} {\bibinfo {author} {\bibfnamefont {P.~Y.}\ \bibnamefont
  {Wen}}, \bibinfo {author} {\bibfnamefont {K.-T.}\ \bibnamefont {Lin}},
  \bibinfo {author} {\bibfnamefont {A.~F.}\ \bibnamefont {Kockum}}, \bibinfo
  {author} {\bibfnamefont {B.}~\bibnamefont {Suri}}, \bibinfo {author}
  {\bibfnamefont {H.}~\bibnamefont {Ian}}, \bibinfo {author} {\bibfnamefont
  {J.~C.}\ \bibnamefont {Chen}}, \bibinfo {author} {\bibfnamefont {S.~Y.}\
  \bibnamefont {Mao}}, \bibinfo {author} {\bibfnamefont {C.~C.}\ \bibnamefont
  {Chiu}}, \bibinfo {author} {\bibfnamefont {P.}~\bibnamefont {Delsing}},
  \bibinfo {author} {\bibfnamefont {F.}~\bibnamefont {Nori}}, \bibinfo {author}
  {\bibfnamefont {G.-D.}\ \bibnamefont {Lin}},\ and\ \bibinfo {author}
  {\bibfnamefont {I.-C.}\ \bibnamefont {Hoi}},\ }\bibfield  {title} {\bibinfo
  {title} {Large collective lamb shift of two distant superconducting
  artificial atoms},\ }\href {https://doi.org/10.1103/PhysRevLett.123.233602}
  {\bibfield  {journal} {\bibinfo  {journal} {Phys. Rev. Lett.}\ }\textbf
  {\bibinfo {volume} {123}},\ \bibinfo {pages} {233602} (\bibinfo {year}
  {2019})}\BibitemShut {NoStop}%
\bibitem [{\citenamefont {Heeg}\ and\ \citenamefont
  {Evers}(2013)}]{heegXrayQuantumOptics2013}%
  \BibitemOpen
  \bibfield  {author} {\bibinfo {author} {\bibfnamefont {K.~P.}\ \bibnamefont
  {Heeg}}\ and\ \bibinfo {author} {\bibfnamefont {J.}~\bibnamefont {Evers}},\
  }\bibfield  {title} {\bibinfo {title} {x-ray quantum optics with
  m{\"o}ssbauer nuclei embedded in thin-film cavities},\ }\href
  {https://doi.org/10.1103/PhysRevA.88.043828} {\bibfield  {journal} {\bibinfo
  {journal} {Phys. Rev. A}\ }\textbf {\bibinfo {volume} {88}},\ \bibinfo
  {pages} {043828} (\bibinfo {year} {2013})}\BibitemShut {NoStop}%
\bibitem [{\citenamefont {Heeg}()}]{pynuss}%
  \BibitemOpen
  \bibfield  {author} {\bibinfo {author} {\bibfnamefont {K.~P.}\ \bibnamefont
  {Heeg}},\ }\href@noop {} {}\bibinfo {howpublished} {Private
  communication}\BibitemShut {NoStop}%
\bibitem [{\citenamefont {Liu}\ \emph {et~al.}(2016)\citenamefont {Liu},
  \citenamefont {Zhao}, \citenamefont {Meng}, \citenamefont {Peng},
  \citenamefont {Dearden}, \citenamefont {Huo}, \citenamefont {Yang},
  \citenamefont {Li},\ and\ \citenamefont {Wen}}]{liu_mossbauer_2016}%
  \BibitemOpen
  \bibfield  {author} {\bibinfo {author} {\bibfnamefont {X.-W.}\ \bibnamefont
  {Liu}}, \bibinfo {author} {\bibfnamefont {S.}~\bibnamefont {Zhao}}, \bibinfo
  {author} {\bibfnamefont {Y.}~\bibnamefont {Meng}}, \bibinfo {author}
  {\bibfnamefont {Q.}~\bibnamefont {Peng}}, \bibinfo {author} {\bibfnamefont
  {A.~K.}\ \bibnamefont {Dearden}}, \bibinfo {author} {\bibfnamefont {C.-F.}\
  \bibnamefont {Huo}}, \bibinfo {author} {\bibfnamefont {Y.}~\bibnamefont
  {Yang}}, \bibinfo {author} {\bibfnamefont {Y.-W.}\ \bibnamefont {Li}},\ and\
  \bibinfo {author} {\bibfnamefont {X.-D.}\ \bibnamefont {Wen}},\ }\bibfield
  {title} {\bibinfo {title} {Mössbauer {Spectroscopy} of {Iron} {Carbides}:
  {From} {Prediction} to {Experimental} {Confirmation}},\ }\href
  {https://doi.org/10.1038/srep26184} {\bibfield  {journal} {\bibinfo
  {journal} {Scientific Reports}\ }\textbf {\bibinfo {volume} {6}},\ \bibinfo
  {pages} {26184} (\bibinfo {year} {2016})}\BibitemShut {NoStop}%
\bibitem [{\citenamefont {Buhmann}(2012)}]{buhmannDispersionForces2012}%
  \BibitemOpen
  \bibfield  {author} {\bibinfo {author} {\bibfnamefont {S.~Y.}\ \bibnamefont
  {Buhmann}},\ }\href {https://doi.org/10.1007/978-3-642-32484-0} {\emph
  {\bibinfo {title} {Dispersion {{Forces I}}}}},\ \bibinfo {series} {Springer
  {{Tracts}} in {{Modern Physics}}}, Vol.\ \bibinfo {volume} {247, Sec. 2.1.4}\
  (\bibinfo  {publisher} {{Springer, Berlin, Heidelberg}},\ \bibinfo {year}
  {2012})\BibitemShut {NoStop}%
\bibitem [{\citenamefont {Jackson}(1999)}]{Jackson}%
  \BibitemOpen
  \bibfield  {author} {\bibinfo {author} {\bibfnamefont {J.~D.}\ \bibnamefont
  {Jackson}},\ }\href@noop {} {\emph {\bibinfo {title} {Classical
  Electrodynamics}}}\ (\bibinfo  {publisher} {{Wiley, New York}},\ \bibinfo
  {year} {1999})\BibitemShut {NoStop}%
\bibitem [{\citenamefont {Hannon}\ and\ \citenamefont
  {Trammell}(1969)}]{hannonMossbauerDiffractionII1969}%
  \BibitemOpen
  \bibfield  {author} {\bibinfo {author} {\bibfnamefont {J.~P.}\ \bibnamefont
  {Hannon}}\ and\ \bibinfo {author} {\bibfnamefont {G.~T.}\ \bibnamefont
  {Trammell}},\ }\bibfield  {title} {\bibinfo {title} {M\"ossbauer
  {{Diffraction}}. {{II}}. {{Dynamical Theory}} of {{M\"ossbauer Optics}}},\
  }\href {https://doi.org/10.1103/PhysRev.186.306} {\bibfield  {journal}
  {\bibinfo  {journal} {Physical Review}\ }\textbf {\bibinfo {volume} {186}},\
  \bibinfo {pages} {306} (\bibinfo {year} {1969})}\BibitemShut {NoStop}%
\bibitem [{\citenamefont {Hannon}\ \emph {et~al.}(1974)\citenamefont {Hannon},
  \citenamefont {Carron},\ and\ \citenamefont
  {Trammell}}]{hannonMossbauerDiffractionIII1974}%
  \BibitemOpen
  \bibfield  {author} {\bibinfo {author} {\bibfnamefont {J.~P.}\ \bibnamefont
  {Hannon}}, \bibinfo {author} {\bibfnamefont {N.~J.}\ \bibnamefont {Carron}},\
  and\ \bibinfo {author} {\bibfnamefont {G.~T.}\ \bibnamefont {Trammell}},\
  }\bibfield  {title} {\bibinfo {title} {M\"ossbauer diffraction. {{III}}.
  {{Emission}} of {{M\"ossbauer}} gamma rays from crystals. {{A}}. {{General}}
  theory},\ }\href {https://doi.org/10.1103/PhysRevB.9.2791} {\bibfield
  {journal} {\bibinfo  {journal} {Phys. Rev. B}\ }\textbf {\bibinfo {volume}
  {9}},\ \bibinfo {pages} {2791} (\bibinfo {year} {1974})}\BibitemShut
  {NoStop}%
\bibitem [{\citenamefont {Hannon}\ and\ \citenamefont
  {Trammell}(1968)}]{hannonMossbauerDiffractionQuantum1968}%
  \BibitemOpen
  \bibfield  {author} {\bibinfo {author} {\bibfnamefont {J.~P.}\ \bibnamefont
  {Hannon}}\ and\ \bibinfo {author} {\bibfnamefont {G.~T.}\ \bibnamefont
  {Trammell}},\ }\bibfield  {title} {\bibinfo {title} {M\"ossbauer
  {{Diffraction}}. {{I}}. {{Quantum Theory}} of {{Gamma}}-{{Ray}} and
  {{X}}-{{Ray Optics}}},\ }\href {https://doi.org/10.1103/PhysRev.169.315}
  {\bibfield  {journal} {\bibinfo  {journal} {Physical Review}\ }\textbf
  {\bibinfo {volume} {169}},\ \bibinfo {pages} {315} (\bibinfo {year}
  {1968})}\BibitemShut {NoStop}%
\bibitem [{\citenamefont {Hannon}\ \emph
  {et~al.}(1985{\natexlab{a}})\citenamefont {Hannon}, \citenamefont {Trammell},
  \citenamefont {Mueller}, \citenamefont {Gerdau}, \citenamefont {R{\"u}ffer},\
  and\ \citenamefont
  {Winkler}}]{hannonGrazingincidenceAntireflectionFilms1985a}%
  \BibitemOpen
  \bibfield  {author} {\bibinfo {author} {\bibfnamefont {J.~P.}\ \bibnamefont
  {Hannon}}, \bibinfo {author} {\bibfnamefont {G.~T.}\ \bibnamefont
  {Trammell}}, \bibinfo {author} {\bibfnamefont {M.}~\bibnamefont {Mueller}},
  \bibinfo {author} {\bibfnamefont {E.}~\bibnamefont {Gerdau}}, \bibinfo
  {author} {\bibfnamefont {R.}~\bibnamefont {R{\"u}ffer}},\ and\ \bibinfo
  {author} {\bibfnamefont {H.}~\bibnamefont {Winkler}},\ }\bibfield  {title}
  {\bibinfo {title} {Grazing-incidence antireflection films. {{III}}.
  {{General}} theory for pure nuclear reflections},\ }\href
  {https://doi.org/10.1103/PhysRevB.32.6363} {\bibfield  {journal} {\bibinfo
  {journal} {Physical Review B}\ }\textbf {\bibinfo {volume} {32}},\ \bibinfo
  {pages} {6363} (\bibinfo {year} {1985}{\natexlab{a}})}\BibitemShut {NoStop}%
\bibitem [{\citenamefont {Hannon}\ \emph
  {et~al.}(1985{\natexlab{b}})\citenamefont {Hannon}, \citenamefont {Trammell},
  \citenamefont {Mueller}, \citenamefont {Gerdau}, \citenamefont {R{\"u}ffer},\
  and\ \citenamefont
  {Winkler}}]{hannonGrazingincidenceAntireflectionFilms1985}%
  \BibitemOpen
  \bibfield  {author} {\bibinfo {author} {\bibfnamefont {J.~P.}\ \bibnamefont
  {Hannon}}, \bibinfo {author} {\bibfnamefont {G.~T.}\ \bibnamefont
  {Trammell}}, \bibinfo {author} {\bibfnamefont {M.}~\bibnamefont {Mueller}},
  \bibinfo {author} {\bibfnamefont {E.}~\bibnamefont {Gerdau}}, \bibinfo
  {author} {\bibfnamefont {R.}~\bibnamefont {R{\"u}ffer}},\ and\ \bibinfo
  {author} {\bibfnamefont {H.}~\bibnamefont {Winkler}},\ }\bibfield  {title}
  {\bibinfo {title} {Grazing-incidence antireflection films. {{IV}}.
  {{Application}} to {{M\"ossbauer}} filtering of synchrotron radiation},\
  }\href {https://doi.org/10.1103/PhysRevB.32.6374} {\bibfield  {journal}
  {\bibinfo  {journal} {Physical Review B}\ }\textbf {\bibinfo {volume} {32}},\
  \bibinfo {pages} {6374} (\bibinfo {year} {1985}{\natexlab{b}})}\BibitemShut
  {NoStop}%
\bibitem [{\citenamefont
  {Sturhahn}(2004)}]{sturhahnNuclearResonantSpectroscopy2004}%
  \BibitemOpen
  \bibfield  {author} {\bibinfo {author} {\bibfnamefont {W.}~\bibnamefont
  {Sturhahn}},\ }\bibfield  {title} {\bibinfo {title} {Nuclear resonant
  spectroscopy},\ }\href {https://doi.org/10.1088/0953-8984/16/5/009}
  {\bibfield  {journal} {\bibinfo  {journal} {J. Phys.: Condens. Matter}\
  }\textbf {\bibinfo {volume} {16}},\ \bibinfo {pages} {S497} (\bibinfo {year}
  {2004})}\BibitemShut {NoStop}%
\bibitem [{\citenamefont {Sturhahn}\ and\ \citenamefont
  {Kohn}(1999)}]{sturhahnTheoreticalAspectsIncoherent1999}%
  \BibitemOpen
  \bibfield  {author} {\bibinfo {author} {\bibfnamefont {W.}~\bibnamefont
  {Sturhahn}}\ and\ \bibinfo {author} {\bibfnamefont {V.}~\bibnamefont
  {Kohn}},\ }\bibfield  {title} {\bibinfo {title} {Theoretical aspects of
  incoherent nuclear resonant scattering},\ }\href
  {https://doi.org/10.1023/A:1017071806895} {\bibfield  {journal} {\bibinfo
  {journal} {Hyperfine Interactions}\ }\textbf {\bibinfo {volume} {123}},\
  \bibinfo {pages} {367} (\bibinfo {year} {1999})}\BibitemShut {NoStop}%
\bibitem [{\citenamefont {Toma{\v
  s}}(1995)}]{tomasGreenFunctionMultilayers1995a}%
  \BibitemOpen
  \bibfield  {author} {\bibinfo {author} {\bibfnamefont {M.~S.}\ \bibnamefont
  {Toma{\v s}}},\ }\bibfield  {title} {\bibinfo {title} {Green function for
  multilayers: {{Light}} scattering in planar cavities},\ }\href
  {https://doi.org/10.1103/PhysRevA.51.2545} {\bibfield  {journal} {\bibinfo
  {journal} {Physical Review A}\ }\textbf {\bibinfo {volume} {51}},\ \bibinfo
  {pages} {2545} (\bibinfo {year} {1995})}\BibitemShut {NoStop}%
\bibitem [{\citenamefont {Johansson}(2011)}]{Johansson2011}%
  \BibitemOpen
  \bibfield  {author} {\bibinfo {author} {\bibfnamefont {P.}~\bibnamefont
  {Johansson}},\ }\bibfield  {title} {\bibinfo {title} {Electromagnetic green's
  function for layered systems: Applications to nanohole interactions in thin
  metal films},\ }\href {https://doi.org/10.1103/PhysRevB.83.195408} {\bibfield
   {journal} {\bibinfo  {journal} {Phys. Rev. B}\ }\textbf {\bibinfo {volume}
  {83}},\ \bibinfo {pages} {195408} (\bibinfo {year} {2011})}\BibitemShut
  {NoStop}%
\bibitem [{\citenamefont {Sharon}\ and\ \citenamefont
  {Tsuei}(1972)}]{sharonMagnetismAmorphousFePdP1972}%
  \BibitemOpen
  \bibfield  {author} {\bibinfo {author} {\bibfnamefont {T.~E.}\ \bibnamefont
  {Sharon}}\ and\ \bibinfo {author} {\bibfnamefont {C.~C.}\ \bibnamefont
  {Tsuei}},\ }\bibfield  {title} {\bibinfo {title} {Magnetism in {{Amorphous
  Fe}}-{{Pd}}-{{P Alloys}}},\ }\href {https://doi.org/10.1103/PhysRevB.5.1047}
  {\bibfield  {journal} {\bibinfo  {journal} {Phys. Rev. B}\ }\textbf {\bibinfo
  {volume} {5}},\ \bibinfo {pages} {1047} (\bibinfo {year} {1972})}\BibitemShut
  {NoStop}%
\bibitem [{\citenamefont {Van~Diepen}\ and\ \citenamefont
  {Popma}(1976)}]{vandiepenMossbauerEffectMagnetic1976}%
  \BibitemOpen
  \bibfield  {author} {\bibinfo {author} {\bibfnamefont {A.~M.}\ \bibnamefont
  {Van~Diepen}}\ and\ \bibinfo {author} {\bibfnamefont {T.~J.~A.}\ \bibnamefont
  {Popma}},\ }\bibfield  {title} {\bibinfo {title} {M{\"o}ssbauer effect and
  magnetic properites of an amorphous {{Fe2O3}}},\ }\href
  {https://doi.org/10.1051/jphyscol:19766158} {\bibfield  {journal} {\bibinfo
  {journal} {J. Phys. Colloques}\ }\textbf {\bibinfo {volume} {37}},\ \bibinfo
  {pages} {755} (\bibinfo {year} {1976})}\BibitemShut {NoStop}%
\bibitem [{\citenamefont {F{\'e}rey}\ \emph {et~al.}(1979)\citenamefont
  {F{\'e}rey}, \citenamefont {Leclerc}, \citenamefont {{de Pape}},
  \citenamefont {Mariot},\ and\ \citenamefont
  {Varret}}]{fereyCaracterisationVarieteAmorphie1979}%
  \BibitemOpen
  \bibfield  {author} {\bibinfo {author} {\bibfnamefont {G.}~\bibnamefont
  {F{\'e}rey}}, \bibinfo {author} {\bibfnamefont {A.}~\bibnamefont {Leclerc}},
  \bibinfo {author} {\bibfnamefont {R.}~\bibnamefont {{de Pape}}}, \bibinfo
  {author} {\bibfnamefont {J.}~\bibnamefont {Mariot}},\ and\ \bibinfo {author}
  {\bibfnamefont {F.}~\bibnamefont {Varret}},\ }\bibfield  {title} {\bibinfo
  {title} {{Caracterisation d'une variete amorphie de FeF3 : Etude thermique,
  magnetique et Mossbauer}},\ }\href
  {https://doi.org/10.1016/0038-1098(79)90789-0} {\bibfield  {journal}
  {\bibinfo  {journal} {Solid State Communications}\ }\textbf {\bibinfo
  {volume} {29}},\ \bibinfo {pages} {477} (\bibinfo {year} {1979})}\BibitemShut
  {NoStop}%
\bibitem [{{\relax DLMF}()}]{NIST:DLMF}%
  \BibitemOpen
  {\relax DLMF},\ \href {http://dlmf.nist.gov/} {\bibinfo {title} {{\it NIST
  Digital Library of Mathematical Functions}}},\ \bibinfo {howpublished}
  {http://dlmf.nist.gov/, Release 1.0.28 of 2020-09-15},\ \bibinfo {note}
  {f.~W.~J. Olver, A.~B. {Olde Daalhuis}, D.~W. Lozier, B.~I. Schneider, R.~F.
  Boisvert, C.~W. Clark, B.~R. Miller, B.~V. Saunders, H.~S. Cohl, and M.~A.
  McClain, eds.}\BibitemShut {Stop}%
\bibitem [{\citenamefont {Fleischhauer}\ \emph {et~al.}(2005)\citenamefont
  {Fleischhauer}, \citenamefont {Imamoglu},\ and\ \citenamefont
  {Marangos}}]{EITReview2005}%
  \BibitemOpen
  \bibfield  {author} {\bibinfo {author} {\bibfnamefont {M.}~\bibnamefont
  {Fleischhauer}}, \bibinfo {author} {\bibfnamefont {A.}~\bibnamefont
  {Imamoglu}},\ and\ \bibinfo {author} {\bibfnamefont {J.~P.}\ \bibnamefont
  {Marangos}},\ }\bibfield  {title} {\bibinfo {title} {Electromagnetically
  induced transparency: Optics in coherent media},\ }\href@noop {} {\bibfield
  {journal} {\bibinfo  {journal} {Rev. Mod. Phys.}\ }\textbf {\bibinfo {volume}
  {77}},\ \bibinfo {pages} {633} (\bibinfo {year} {2005})}\BibitemShut
  {NoStop}%
\bibitem [{\citenamefont {Ressayre}\ and\ \citenamefont
  {Tallet}(1977)}]{ressayreQuantumTheory1977}%
  \BibitemOpen
  \bibfield  {author} {\bibinfo {author} {\bibfnamefont {E.}~\bibnamefont
  {Ressayre}}\ and\ \bibinfo {author} {\bibfnamefont {A.}~\bibnamefont
  {Tallet}},\ }\bibfield  {title} {\bibinfo {title} {Quantum theory for
  superradiance},\ }\href {https://doi.org/10.1103/PhysRevA.15.2410} {\bibfield
   {journal} {\bibinfo  {journal} {Phys. Rev. A}\ }\textbf {\bibinfo {volume}
  {15}},\ \bibinfo {pages} {2410} (\bibinfo {year} {1977})}\BibitemShut
  {NoStop}%
\end{thebibliography}%

\appendix

\section{Approximating single particle decay as collective}\label{app:collective-decay}
To justify approximating the single particle decay as collective, we note that we can write the exact form of $L_{\text{A}}$ as
\begin{widetext}
\begin{equation}
\begin{aligned}
 L_A[\rho] &= - \sum_{\mu,i} \hbar\gamma_\mu
 \left(  
 	\ketbra{e_\mu^{(i)}}\rho + \rho\ketbra{e_\mu^{(i)}}
 - 2 \ketbra{g_\mu^{(i)}}{e_\mu^{(i)}}\rho\ketbra{e_\mu^{(i)}}{g_\mu^{(i)}}
  \right)
  \\
  &= -\sum_\mu \hbar\gamma_\mu \left(
  b_\mu^\dagger b_\mu \rho + \rho b_\mu^\dagger b_\mu + \sum_i \ketbra{g_\mu^{(i)}} \mel{e_\mu^{(i)}}{\rho}{e_\mu^{(i)}}
  \right).
\end{aligned}
\end{equation}
\end{widetext}
At low saturations $\mel{e_\mu^{(i)}}{\rho}{e_\mu^{(i)}}\approx 0$, therefore both the single particle and collective decays give the same result,
\begin{equation}
	L_A[\rho] \approx -\sum_\mu \hbar \gamma_\mu \left\{b_\mu^\dagger b_\mu, \rho\right\}.
\end{equation}
In addition, for x-ray quantum optics the cavities considered are in the so-called ``bad cavity'' regime, and the cavity mode lifetime is much shorter than the natural lifetime of the resonant transitions. Therefore, the incoherent processes are dominated by collective interactions, and the internal decay can be approximated as a small correction to the collective decay rates.

\section{Multiple ground states}\label{app:multiple-ground}
We now consider an atom with multiple ground states as well as multiple excited states. As with the excited states, the ground states are considered to have small splittings compared with the energy difference between the ground and excited bands, and dipole transitions are forbidden between ground states. We index here the excited states with Greek indices $\mu,\nu\ldots$, and the ground states with Latin indices $j\ldots$.

For nuclei with multiple ground states, we further partition the ensembles by the initial ground states of the atoms, creating permutationally invariant sub-ensembles. For a given sub-ensemble, we use the initial ground state as the `vacuum' state for the generalised Holstein-Primakoff transformation. We then obtain
\begin{equation}
\begin{aligned}
	\ketbra{\mu}{0}  &\approx \sqrt{N}b_\mu^\dagger,
	\\
	\ketbra{j}{0} &\approx \sqrt{N} c_j^\dagger,
	\\
	\ketbra{j}{k} &= c_j^\dagger c_{k},
	\\
	\ketbra{\mu}{j} &= b_\mu^\dagger c_j,
	\\
	\ketbra{\mu}{\nu} &= b_\mu^\dagger b_\nu,
	\\
	[b_\mu, b_\nu^\dagger] &= \delta_{\mu\nu},
	\\
	[c_j, c_k^\dagger] &= \delta_{jk},
	\\
	[c_j, b_\mu^\dagger] &= 0.
\end{aligned}
\end{equation}
The single particle Hamiltonian then reads
\begin{equation}
	H_{\text{A}} = \sum_{\mu} (\Delta_\mu - \delta_0) b_\mu^{\dagger}b_\mu  + \sum_{j} (\delta_j - \delta_0) c_j^\dagger c_j
\end{equation}
with the single particle decay reading
\begin{equation}
	L_{\text{A}}[\rho] = -\sum_{\mu}\gamma_{\mu 0}\mathcal{L}[\rho, b_\mu^\dagger, b_\mu]
	-\sum_{\mu,j}\gamma_{\mu j}\mathcal{L}[\rho, b_\mu^\dagger c_j, c_j^\dagger b_\mu],
\end{equation}
where $\gamma_{\mu 0}$ is the decay rate to the initial ground state, and $\gamma_{\mu j}$ is the decay rate to ground state $j$. If we consider the action of this super-operator on $b_\mu$, we obtain
\begin{equation}
	L_{\text{A}}[b_\mu] = -\gamma_{\mu 0} b_\mu - \sum_j \gamma_{\mu_j} (1 + c_j^\dagger c_j)b_\mu.
\end{equation}
The action on a ground state operator reads
\begin{equation}
	L_A[c_j]=\sum_\mu \gamma_{\mu j} b_\mu^\dagger b_\mu c_j.
\end{equation}
The rate of population transfer to the other ground state is thus proportional to the excited state populations, which are negligible in the linear response regime. Therefore, we can assume $c_j c_k \approx 0$ for all $j,k$. However, the decay rate of the transition operator $b_\mu$ is still affected by the left over terms,
\begin{equation}
L_{\text{A}}[b_\mu] \approx -(\gamma_{\mu 0}+ \sum_{j} \gamma_{\mu j}) b_\mu = -\gamma_\mu b_\mu,
\end{equation}
i.e, the effective decay rate $\gamma_\mu$ for a transition operator $b_\mu$ is the sum of the decay rates of all decay channels for excited state $\mu$. Thus, we may write
\begin{equation}
L_{\text{A}}[\rho] \approx -\sum_\mu \gamma_\mu \mathcal{L}[\rho, b_\mu^\dagger, b_\mu].
\end{equation}

In addition to the internal decay, the cavity mediated coupling includes transitions to different ground states than the initial. The interaction Hamiltonian reads 
\begin{equation}
\begin{aligned}
	H_{\text{dd}} = 
	&-J \sum_{\mu,\nu} \vb{d}_{\mu 0}^\perp\cdot \vb{d}_{\nu 0}^{*\perp} b_\mu^\dagger b_\nu 
	\\
	&-\frac{J}{\sqrt{N}}\sum_{\mu,\nu}\sum_{m} \vb{d}_{\mu0}^\perp \cdot \vb{d}_{\nu j}^{*\perp} b_\mu^\dagger b_\nu c_j^\dagger + \mathrm{h.c.}
	\\ 
	&- \frac{J}{N}\sum_{\mu,\nu}\sum_{j,k} b_\mu^\dagger b_\nu c_j^\dagger c_k.
\end{aligned}
\end{equation}
We can see that terms involving population transfer between ground states are suppressed by a factor of $\sqrt{N}$ or higher, and thus can be neglected in the linear response regime. An analogous argument holds for the dipole-dipole Lindblad term, and for the probe field driving. 

Thus, we conclude that for the linear response in the presence of multiple ground states, only the initial ground state of a sub-ensemble needs to be considered, with the only contributions of the other ground states being to the decay rates of the transition operators.

\end{document}